# Replicating the behaviour of electric vehicle drivers using an agent-based reinforcement learning model

Zixin Feng, Qunshan Zhao, Alison Heppenstall
Urban Big Data Centre, School of Social and Political Sciences, University of Glasgow, United Kingdom, G12 8RZ

## Abstract

Despite the rapid expansion of electric vehicle (EV) charging networks, questions remain about their efficiency in meeting the growing needs of EV drivers. Previous simulation-based approaches, which rely on static behavioural rules, have struggled to capture the adaptive behaviours of human drivers. Although reinforcement learning has been introduced in EV simulation studies, its application has primarily focused on optimising fleet operations rather than modelling private drivers who make independent charging decisions. To address the gap, we propose a multi-stage reinforcement learning framework that simulates charging demand of private EV drivers across a national-scale road network. We validate the model against real-world data and identify the training stage that most closely reflects actual driver behaviour, which captures both the adaptive behaviours and bounded rationality of private drivers. Based on the simulation results, we also identify critical 'charging deserts' where EV drivers consistently have low state of charge. Our findings also highlight recent policy shifts toward expanding rapid charging hubs along motorway corridors and city boundaries to meet the demand from long-distance trips.

# 1 Introduction

In the past few years, with global net-zero carbon transition targets, rapid technological advancements, and declining battery costs, many countries have witnessed a significant rise in EV adoption (Mahmud et al., 2023). In the early stages of EV adoption, households that adopt EVs typically own garages or off-street parking suitable for installing home chargers (LaMonaca and Ryan, 2022). However, as EV ownership continues to grow, EV adoption has shifted from early adopters to the mass market (IEA, 2024). This transition has led to increased demand for public chargers, similar to the ubiquitous gas stations. This demand is particularly high among drivers without access to private chargers, such as those living in flats.

Recent research has questioned whether the emission reduction benefits from EVs can be sustained due to variability in consumer behaviour (Nunes et al., 2022). Concerns are also raised about whether the charging network expansion is misallocating resources and failing to adapt to user needs (Metais et al., 2022). In the UK, geographical disparities in public chargers have created significant barriers to EV adoption. Some areas have lagged behind others in the deployment process (Peng et al., 2024). Additionally, survey by Department for Transport (2022) shows that existing EV drivers aim to adjust their behaviours to integrate charging into their trip schedules and daily habits. Meanwhile, potential EV adopters prioritise proximity, reliability and dependability of chargers as key factors in their decision to switch to EVs. It is therefore important to understand the complex behaviours of diverse EV drivers and accurately estimate charging demand to support a more effective and equitable charging network.

Previous studies have explored EV driver behaviours and estimated their charging demand using either data-driven or simulation-based approaches. Many studies are conducted within grid cells (van der Kam et al., 2019) or intra-city scale networks (Pagani et al., 2019; Yi et al., 2023), while other studies also extend to large geographical contexts large geographical contexts over regional or country levels where long-distance journeys bring critical challenges to EV drivers (Anjos et al., 2020; Liao et al., 2023). However, despite their contributions, both approaches have limitations. Data-driven methods often face limited access to EV-specific datasets due to privacy concerns (Park and Joe, 2024). As a result, researchers often rely on alternative datasets, such as socio-demographic data (Crozier et al., 2021), GPS data from conventional vehicles (Kontou et al., 2019) or commercial EV fleets (Hu et al., 2022). While these datasets offer high granularity or rich attributes, they may lead to biased representations of private EV drivers. Meanwhile, simulation-based approaches, such as Agent-based Models (ABMs), have proven effective in capturing the unique features of EV drivers (Feng et al., 2025; Yi et al., 2023). However, these models often rely on static behavioural rules, which fail to reflect the adaptive nature of human drivers in response to the changing environment. For example, as EV drivers gain experience, they tend to charge less frequently and become more comfortable operating with lower battery levels. Lastly, a critical gap in most simulation-based studies is the lack of robust validation against real-world data, which undermines the credibility of their findings.

The objective of this research is to develop and validate an agent-based reinforcement learning (RL) model using a Deep-Q Network (DQN) to (1) simulate the adaptive charging behaviours of heterogeneous EV drivers in Great Britain (GB); and (2) estimate the spatial distribution of public charging demand and identify locations with high risks of battery depletion. Our approach first categorises drivers based on key characteristics, including travel distance, battery status, trip purpose, and environmental factors. We then implement a novel multi-episode training-simulation framework where representative agents from each cluster learn charging behaviours through RL. These learned behaviours are then used to simulate the broader population of drivers within their respective groups, creating an updated charging environment that feeds back into the next training episode. This iterative process captures both the adaptive charging behaviours of EV drivers and the changing interactions of drivers at charging stations. More importantly, our model differs from previous studies which use RL with rational agents to derive optimal routing and charging strategies (Dastpak et al., 2024; Jin and Xu, 2022; Yu et al., 2023). We use RL to replicate the adaptive learning process of real-world private drivers operating under bounded rationality when making charging decisions. Rather than identifying the training episode that achieves optimal performance, we validate against real-world charging session data to identify the episode that most closely reflects observed behaviour. Additionally, it provides a validated modelling framework to estimate charging demand distribution and identify locations that need new public chargers. It can contribute to a more accurate charging demand estimation of private EV drivers and informing future charging infrastructure planning.

# 2 Background

To date, there have been extensive studies on charging demand estimation and EV driver behaviour, primarily based on data-driven or simulation-based approaches. Data-driven studies often rely on socio-demographic data, such as population density, average travel distance, and travel flow volumes, to estimate charging demand distribution (Dong et al., 2019; Hardinghaus et al., 2019). More recent data-driven research has also used vehicle trajectories to infer EV charging demand, including data from conventional vehicles (Kontou et al., 2019) and EV fleets like taxis (Hu et al., 2022). While these data offer high granularity, they are still controversial in terms of biased representation for private EV drivers. Meanwhile, limited data availability for private EV drivers remains a challenge, partly due to the early adoption stage of EVs and privacy concerns about connected vehicle data. One notable exception is the use of surveys to collect data on private EV drivers (Hasan and Simsekoglu, 2020). While these surveys provide valuable insights into driver behaviours, their high costs often restrict sample sizes and geographical scale, and limit their generalisability to broader spatial contexts.

Given the lack of data for private EV drivers, simulation-based approaches, such as ABMs, have been developed to simulate EV charging demand with predefined behavioural rules. Simulation models consider multiple aspects of driver behaviours, including psychological features, financial preferences, and vehicle attributes (Liu et al., 2022; Adenaw and Lienkamp 2021). However, despite these detailed behavioural rules, questions remain about their ability to represent EV drivers' diversity and adaptability. Dahlke et al. (2020) argues that there is a need for ABMs to better capture the heterogeneous behaviour of agents. However, incorporating diverse agent types requires complex behavioural rules with large parameter spaces, which can pose verification challenges (Sun et al., 2016). Additionally, previous ABMs with static behaviour rules are limited in capturing the adaptive behaviours of EV drivers. Drivers are humans who can learn from past experiences and adjust their behaviours over time. As indicated by Macal (2016), ABMs should integrate adaptability and learning capabilities of the agents to improve the explanatory power and credibility in simulations.

Compared to ABMs, RL has demonstrated advantages in capturing adaptive behaviours of humans, where positive behaviours are learned through repeated exposure to an environment (Malleson et al., 2022). RL is a goal-directed learning framework and is considered a model-free approach for studying sequential and dynamic decision-making problems. Unlike other machine Learning methods, RL is particularly valuable when no dataset of correct answers exists (Dahlke et al., 2020), making it especially useful for studying private EV drivers whose trajectory data is limited.

In transportation research, RL has been widely applied to optimise vehicle routing and charging decisions (Jebessa et al., 2022; Koh et al., 2020). One notable feature of these studies is their emphasis on optimising routing and charging strategies for EVs (Lee et al., 2020; Yan et al., 2021). One group of studies assume agents share information with each other in a timely manner and use RL to optimise coordinated decision-making of EV fleets (Lin et al., 2022; Sultanuddin et al., 2023). Another group of studies focuses on individual driver optimisation, where RL is used to identify optimal routing and charging strategies an EV driver should adopt, in order to minimise travel time, energy consumption, or charging costs (Dastpak et al., 2024; Jin and Xu, 2022; Yu et al., 2023). However, the assumptions of complete information exchange and perfect rationality do not apply to private EV drivers, who have limited, random communication with each other and act independently. Recent research on modelling cognitive decision-makers has increasingly emphasised the importance of incorporating bounded rationality into human decision-making models (Manley and Cheng, 2018; Pappalardo et al., 2023). Despite RL's ability to reflect adaptive learning, there remain gaps in integrating bounded rationality into RL frameworks, which is important for better representing the private EV drivers.

Another limitation of previous studies using RL to train and simulate EV drivers is the challenge of scalability. Exploring EV driver behaviours in large geographical area is critical as it is where long-distance journeys bring critical challenges to EV drivers. There have been studies using ABMs to simulate EV driver behaviours in large geographical scales, such as those by Anjos et al. (2020) and Liao et al. (2023). Nevertheless, incorporating RL to such large-scale environments can create further computational challenges due to the larger road networks and expanded charging networks, which increases the number of charging choices available to agents. While some efforts have been made to improve cooperative multi-agent RL, simulating RL-trained agents in geographically large-scale scenarios involving thousands of independent agents with limited communication with each other needs further exploration.

In summary, despite significant progress in estimating charging demand and analysing EV driver behaviour through both data-driven and simulation-based approaches, several critical gaps remain. Data-driven methods face challenges related to biased representations and limited data availability for private EV drivers. Simulation-based models often rely on static behavioural rules and fail to capture the adaptive and heterogeneous nature of human decision-making. RL offers advantages in incorporating adaptive learning process. However, previous studies mainly use RL for EV fleet optimisation. It remains underexplored about how to use RL to simulate private EV drivers with bounded rationality and limited communications with each other, especially in large-scale geographical contexts where computational burdens increase.

# 3 Data and study area

We take GB as our study area, which comprises England, Scotland, and Wales. We focus on the trips made by England residents travelling across GB. Over the past decade, GB has seen consistent growth in its public charging network. As of April 2024, there are 59,125 publicly available chargers in the region (Department for Transport, 2024). Additionally, England has a high volume of trips taken by private vehicles, with its residents travelling an average of 5,373 miles in 2022 (Department for Transport, 2023a).

In our model, agents represent a combination of EV drivers and their vehicles. Travel schedules of EV drivers are established using attributes derived from National Travel Survey data (Department for Transport, 2023b), combined with Ordnance Survey (OS) Code-Point data (Ordnance Survey, 2023a). EV attributes are sourced from the EV Database (EV Database, 2024). The simulation environment integrates road networks from OS Open Roads dataset (Ordnance Survey, 2023b) and public chargers from ChargePoint dataset (ChargePoint, 2023). The availability status of chargers is scraped from the ChargePoint website and are used for model validation (ChargePoint, 2023). Further details of data and study area are provided in the Data and Study Area section in the Supplementary Materials.

# 4 Methods

### 4.1 Multi-stage Training-Simulation Framework

To address the limitations discussed in Section 2, we propose a scalable multi-stage framework in which, at each training episode, representative agents from each cluster are trained. Their current policies are then used to simulate all agents within their respective clusters. Through continuous validation against empirical data at each episode, the framework identifies the charger usage patterns that most closely reflect real-world situation. This approach provides a practical solution for modelling large-scale systems. It allows for the inclusion of larger agent populations and application across diverse scenarios while maintaining computational efficiency.

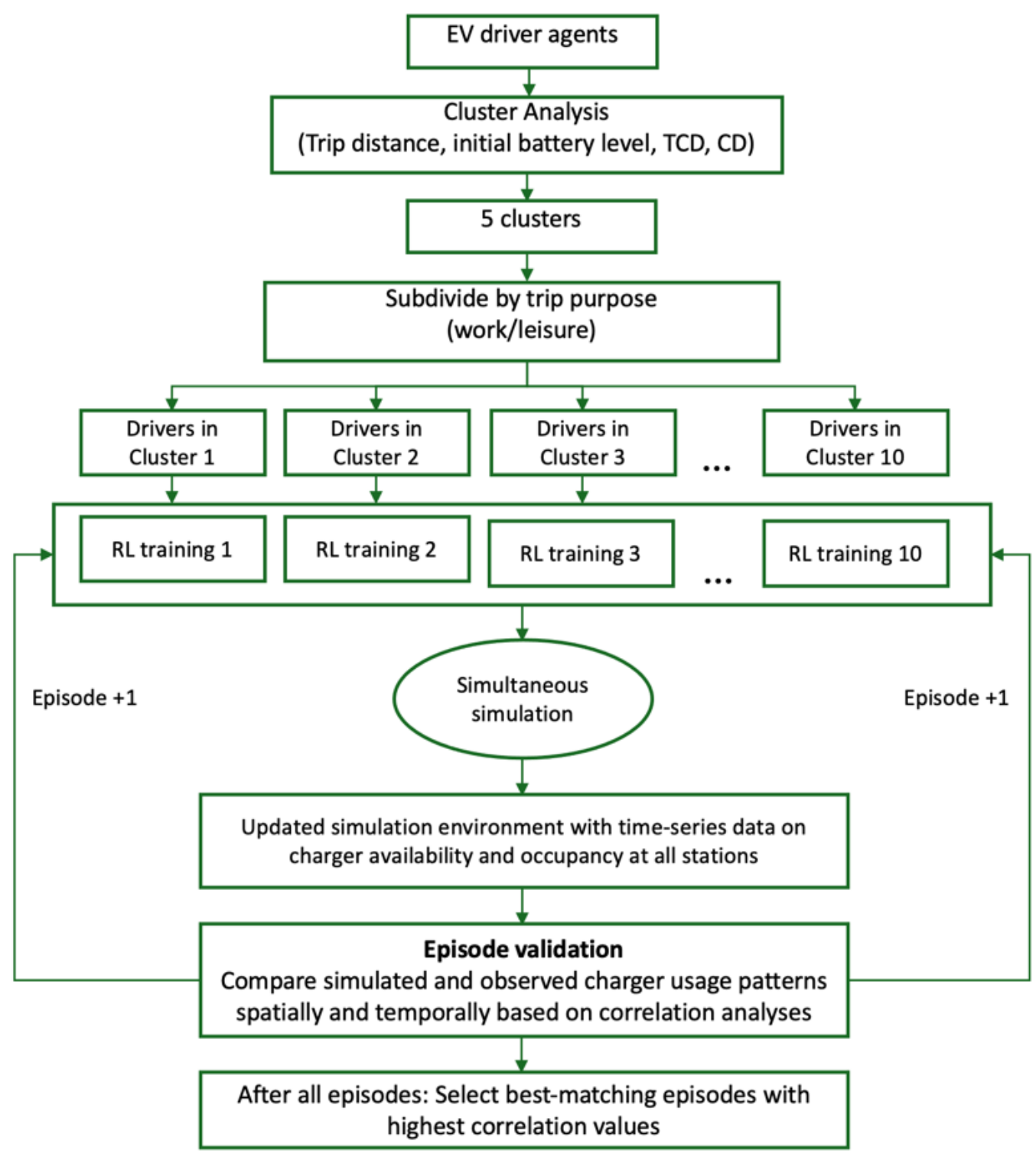

Figure 1. Methodology framework.

The details of the methodology framework are shown in Figure 1. Specifically, we first cluster all EV drivers from the travel survey dataset based on their trip distance, initial battery level, Trip co-occurrence density (TCD), and Charger density (CD). The initial battery level, TCD and CD are not directly provided by the dataset, and the methods of calculating them are detailed in the Data and Study Area section in the Supplementary Materials. Trip purpose (work, leisure) was not directly included in the cluster analysis due to both the limited sample size of leisure trips and the categorical nature of this variable, which could bias the clustering results. Instead, after clustering analysis, each group is further subdivided based on the trip purpose to account for the different behavioural patterns associated with work and leisure. After conducting the clustering analysis, representative agents are selected to initialise the training.

For each cluster, the two agents closest to the cluster centroid are first identified, as these agents best represent the typical characteristics of their respective clusters. These two agents are then assigned to two separate training sets, ensuring that each training set contains one representative agent per cluster. Meanwhile, simulation agents are independently sampled from the full agent population. Ten simulation agent sets are generated, each containing a different random sample of agents. To ensure the robustness of the model and to minimise data overfitting, the model is run for all combinations of the two training sets and the ten simulation sets, resulting in a total of twenty model runs. This number of runs was chosen to balance statistical robustness with computational resource constraints.

Details of the RL model is specified in the RL Model section in Supplementary Materials. The trained policies from each episode are then fed into the simulation model to guide the behaviour of all agents within their respective groups. The simulation phase brings all agents together in a shared environment where their charging decisions affect charger availability for others. During the simulation process, agents interact with each other in two ways: (1) All EV drivers are assumed to have access to the real-time charger status updates via online apps or websites. When a driver occupies a charger, the status of the charger changes from 'available' to 'in use', and this is updated to all other drivers; (2) When a charger is occupied, this does not necessarily prevent other drivers from heading to the same charger. Due to limited availability of chargers in certain areas, drivers may still choose to travel to an in-use charger and queue behind other vehicles. Charging apps and websites do not display queue lengths, and drivers must estimate potential waiting time based on personal experiences. After each episode, the generated charger usage patterns, including the usage time window of each charger, will be incorporated into the state space for the next training episode. This iterative feedback loop enables agents to learn from system-wide charging patterns based on past experiences. The model is executed on AWS, using 64 vCPUs and 128 GB of RAM. Key parameters, including the learning rate, discount factor, and exploration rate, are calibrated to enhance the model's performance.

At the end of each episode, we validate the model by comparing simulated charger usage patterns, including the spatial distribution and temporal windows of charging sessions, against real-world charging session data from ChargePoint dataset. This comparison is performed using spatial and temporal correlation analysis detailed in Section 5.3. After all episodes are complete, we select the episodes with the highest correlation scores to represent the stage where simulated behaviour most closely aligns with observed real-world patterns. It is important to note that the best-matching episode does not correspond to the final episode where agents have fully converged to optimal behaviour. Instead, the episode with the highest correlation to real-world data typically falls somewhere between the early exploratory stage and the final convergence stage. This intermediate stage captures a realistic mix of driver experience levels within the population. The episode that best matches real-world patterns therefore represents a population-level blend of different experience stages, instead of a uniformly optimal outcome.

### 4.2 RL Model

In this paper, we use off-policy RL – Q-learning to enable human drivers to go beyond the current policy and learn from diverse past experiences. While value-based Q-learning performs well in many scenarios with discrete states and actions, its reliance on tabular approach limits its scalability. Training becomes inefficient when dealing with large state and action spaces, such as those in our study. To address this challenge, we employ the Deep Q-Network (DQN) framework introduced by Mnih et al. (2015). By leveraging deep neural networks to approximate the Q-value function to enhance performance, DQN is a more practical choice for our large-scale EV simulation.

Throughout this process, the agent's objective is to complete all planned trips while avoiding battery depletion. Specifically, we formulate the charging decision-making process of EV drivers as a RL problem, where agents observe a state, stake an action, receive a reward, and undergo a transition based on their chosen action. Each agent is initialised with a State of Charge (SOC) and a daily trip schedule comprising one or more destinations. The agent navigates through road networks using Dijkstra shortest path algorithm, moving from its current node at time $t$ ($P_t$) to the next node at time $t$+1 ($P_{t+1}$) while continuously evaluating charging decisions. At any time step $t$, the agent's state $S_t$ is represented as a tuple of five variables:

$$S_t = (SOC_t,\ Dist_t, Time_t,\ Stations_t, Trip_t) \quad (1)$$

Where $SOC_t$ represents the $SOC$ at time $t$, indicating the vehicle's battery status. $Dist_t$ is the shortest distance from the current position to the destination. $Time_t$ indicates the time in minutes past midnight on the given day at time step $t$. $Trip_t$ is the number of remaining trips in the driver's daily trip plan. $Stations_t$ is the percentage of available stations relative to all reachable stations within the vehicle's remaining driving range, based on its $SOC_t$ and energy consumption rate at location $P_t$ at time $t$. At each updated time step ($t$+1), the agent re-evaluates the reachable stations, and recalculates $Stations_{t+1}$ based on its updated location ($P_{t+1}$) and updated remaining SOC ($SOC_{t+1}$).

At each time step t, the agent selects an action $a_{t,i}$ from the action space $A_t$ using a $\epsilon$ greedy strategy. The action space $A_t$ is defined as follows

$$A_t = \{a_{t,0}\} \cap \{a_{t,n,k} | \ n \ \in \{10, 20,30 \dots 100\}, k \in K\} \quad (2)$$

Where:

- $a_{t,0}$ indicates moving from the current node $P_t$ to $P_{t+1}$ without charging.
- $a_{t,n,k}$ includes rerouting to one of the reachable and available charging stations $k$ ($k \in K$) and charging the battery by n%

We implement the learning process using a Deep Q-Network (DQN), which maps state-action pairs to Q-values through hidden layers. The DQN takes the state variables $S_t$ as input and outputs Q-values for each possible action $a_{t,i}$. The Q-value, $Q(s, a)$, represents the agent's expected cumulative reward from taking action $a$ in state $S$:

$$Q(s, a) = r + \gamma \max_{a'} Q(s', a') \quad (3)$$

$r$ is the immediate reward for the action. $\gamma$ is the discount factor $\max_{a'} Q(s', a')$ represents the maximum estimated reward achievable from the next state $s'$. The action follows an ε-greedy strategy for action selection to balance exploration and exploitation. With a probability of $\epsilon$, the agent explores by randomly selecting an action from the available actions. The exploration rate $\epsilon$ is initialised at 0.99 to encourage exploration during the early stages of training. As training progresses, $\epsilon$ gradually decayed over time from 0.99 and reach a minimum value of 0.01 ($1 - \epsilon$) by the end of training (2000 episodes). The decay ensures that agents transition from predominantly exploratory behaviour to predominantly exploitative behaviour. It ensures the agent learns from diverse experiences while refining its decision-making over time.

After performing the selected action $a_{t,i}$, the agent receives a reward $R_t$ , and transitions to a new state $S_{t+1}$. These transitions are stored as experiences in an experience replay buffer (Equation (4)-(5)). During training, random minibatches of experiences are sampled from the replay buffer $D_t$ during the training. This approach helps break temporal correlations and improves learning efficiency by allowing the agent to revisit rare transitions multiple times, thereby enhancing stability in the learning process ((Mnih et al., 2015; Schaul et al., 2015).

$$e_t = \{S_t, a_{t,i}, R_t, S_{t+1}\} \quad (4)$$

$$D_t = \{e_1, e_2, \dots, e_t\} \quad (5)$$

For each mini-batch, the network computes the predicted Q-values and the target Q-values using the Bellman Equation. The network parameters are then updated by minimising the loss function $L(\theta)$, which is the mean squared error of the target Q -value and predicted Q-value:

$$L(\theta) = \mathbb{E}_{s,a,s',a'}\left[\left(r + \gamma \max_{a'} Q(s',a';\, \theta') - Q(s,a;\, \theta)\right)^2\right] \tag{6}$$

Another important term to define the agent's experience is the reward $R_{\mathrm{t}}$. It is the immediate reward received when taking the action $a_{t,i}$. The reward considers financial costs, charging urgency, trip status, and SOC threshold of drivers (Equations (7) -(11)).

$$\mathrm{D}_{SOC} = SOC_t - SOC_{threshold} \tag{7}$$

$$D_{charge} = \mathbb{I}_{charge,t} \cdot \varepsilon \cdot ln\left(timing_{charge} \cdot chance_{charge} \cdot status_{battery} + 1\right) \tag{8}$$

$$\mathbb{I}_{charge,t} = \begin{cases} 1, & if\ charge\ at\ t \\ 0, & if\ not\ charge\ at\ t \end{cases} \tag{9}$$

$$D_{cost} = \left(\alpha \cdot log\left(1 + Payment \cdot \mathbb{I}_{charge,t}\right) + \beta \cdot log\left(1 + T_{travel} + T_{charge} \cdot \mathbb{I}_{charge,t}\right) + 1\right)^{\frac{n}{m}} \tag{10}$$

$$R_{\mathrm{t}} = \frac{\gamma \cdot D_{SOC}}{D_{charge} \cdot D_{cost}} + \rho \cdot Status \tag{11}$$

Where:

- $\mathrm{D}_{SOC}$ is the SOC threshold impact. It is the difference between the current SOC level at $t$ and the drivers' SOC threshold $SOC_{threshold}$. It measures how well the driver is maintaining a sufficient battery level relative to their psychological threshold.
- $D_{charge}$ is the charging urgency factor. It evaluates the urgency of a charging decision. This variable is only considered if a charge behaviour happens.
- $timing_{charge}$ is the lengths of time between the trips starts and the charging activity starts, indicating whether charging occurs early or late in the trip.
- $chance_{charge}$ is the charging opportunity ratio. It is defined as the ratio of CD to TCD, as described in the Data and Study Area section in the Supplementary Materials.
- $status_{battery}$ is the start battery level of a driver.
- $D_{cost}$ is the cost impact. It consists of the financial payment for charging, the time spent for rerouting to charging station, and the actual charging duration.
- $n$ is the number of times the agent charges. It introduces an exponential penalty when the total charge time increases.
- $m$ is a categorical variable of the trip distance class. Compared with longer-distance trips, shorter-distance trips will incur a greater penalty for repetitive charging.
- $Status$ is a categorical variable with values 0, 1, or -1. It represents the driver's trip completion status. A value of -1 indicates results a penalty and indicates that the driver fails to complete the trip. A value of 0 means the driver continues the trip without additional rewards or penalties. A value of 1 means the driver earns an additional reward by successfully reaching a destination.

- $\rho$ is the weight coefficient for the trip completion status. It determines the magnitude of the reward associated with trip completion outcomes.

# 5 Results

## 5.1 Cluster analysis, reward change, and model validation

The cluster analysis was conducted based on four continuous features: total trip distance, initial battery level, trip co-occurrence density, and public charger density. The results of the cluster analysis are included in Supplementary Materials. The cumulative reward change for each type of training agent across all episodes can be found in Figure S8 in Supplementary Materials.

The model was validated using 21 days of real-world charging session data. Cross-correlation analysis was performed between the simulated results of each episode and the daily charging session data, both spatially and temporally. Details of the model validation results are included in the Supplementary Materials. To further illustrate the behavioural validity of the results, we analysed the action selection probabilities of agents throughout the training process and evaluated if the emergent behaviours of different types of trained agents align with established and intuitive understanding of EV driver behaviour. Details are shown in the Action selection Possibility of Agents section in Supplementary Materials.

## 5.2 Charging demand distribution

After model validation, we analysed the SOC distribution along road networks using data from the 703rd episode, which is one of the episodes with highest spatial and temporal correlation with real-world data. Our model generated SOC values for each network node based on all passing EV drivers, and we calculated the mean SOC at each node to represent the average battery status at that location.

Due to the dense distribution of network nodes, we used the H3 hexagonal spatial indexing system to better visualise SOC changes. H3 is an open-source geospatial grid system that divides the Earth's surface into hexagonal cells at varying resolutions. It provides a consistent and scalable method for spatial aggregation (H3, 2024). Lower resolution levels correspond to larger hexagons and broader spatial aggregation, while higher levels offer finer spatial detail.

We calculated mean SOC values for each hexagonal cell and visualised the results at resolutions 7 and 6 (Figures 2 (a) and (b)). Resolution 7 cells have an average area of approximately 5.2 km², and resolution 6 cells average around 36.1 km².

To identify areas where EV drivers face a higher risk of battery depletion, we classify the resolution 6 SOC map into five categories using natural breaks. We focus on locations where SOC falls into the lowest category (0 - 0.22) (Figure 2 (c)). However, these patterns alone do not definitively indicate charging risk, as they may simply reflect drivers' willingness to operate at lower battery levels rather than highlighting areas where charging infrastructure is insufficient. To accurately identify true risk areas, it is essential to consider both SOC patterns and the availability of public chargers together.

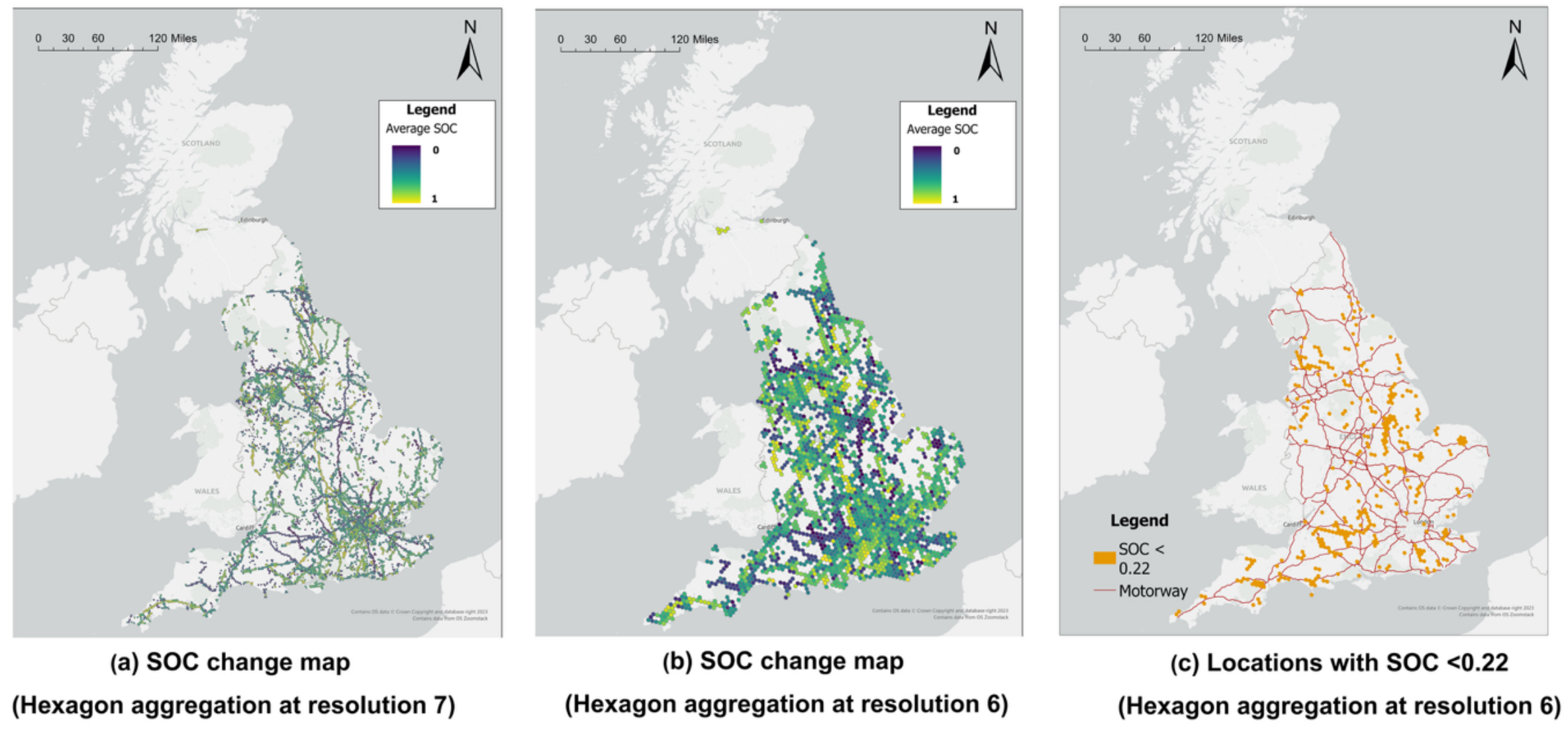

(a) SOC change map (Hexagon aggregation at resolution 7)

(b) SOC change map (Hexagon aggregation at resolution 6)

(c) Locations with SOC <0.22 (Hexagon aggregation at resolution 6)

Figure 2. Consecutive SOC change distribution along road networks at the 703rd episode.

We therefore conducted this combined analysis by creating buffer zones of 500m, 1000m, and 1500m around each node and calculating the number of chargers within each buffer. Using k-means clustering with the charger counts and the average SOC values, we identified three distinct clusters for each buffer distance (Figures 3(a), (b), and (c)). The specific process of selecting the optimal number of clusters is included in the Cluster Selection section in the Supplementary Materials. Across all buffer distances, Cluster 0 consistently represented locations with both low average SOC values and limited nearby charging infrastructure, which can indicate higher risks of battery depletion for EV drivers. The spatial distribution of these high-risk locations varies with buffer size and are shown in Figures 3(d), (e), and (f).

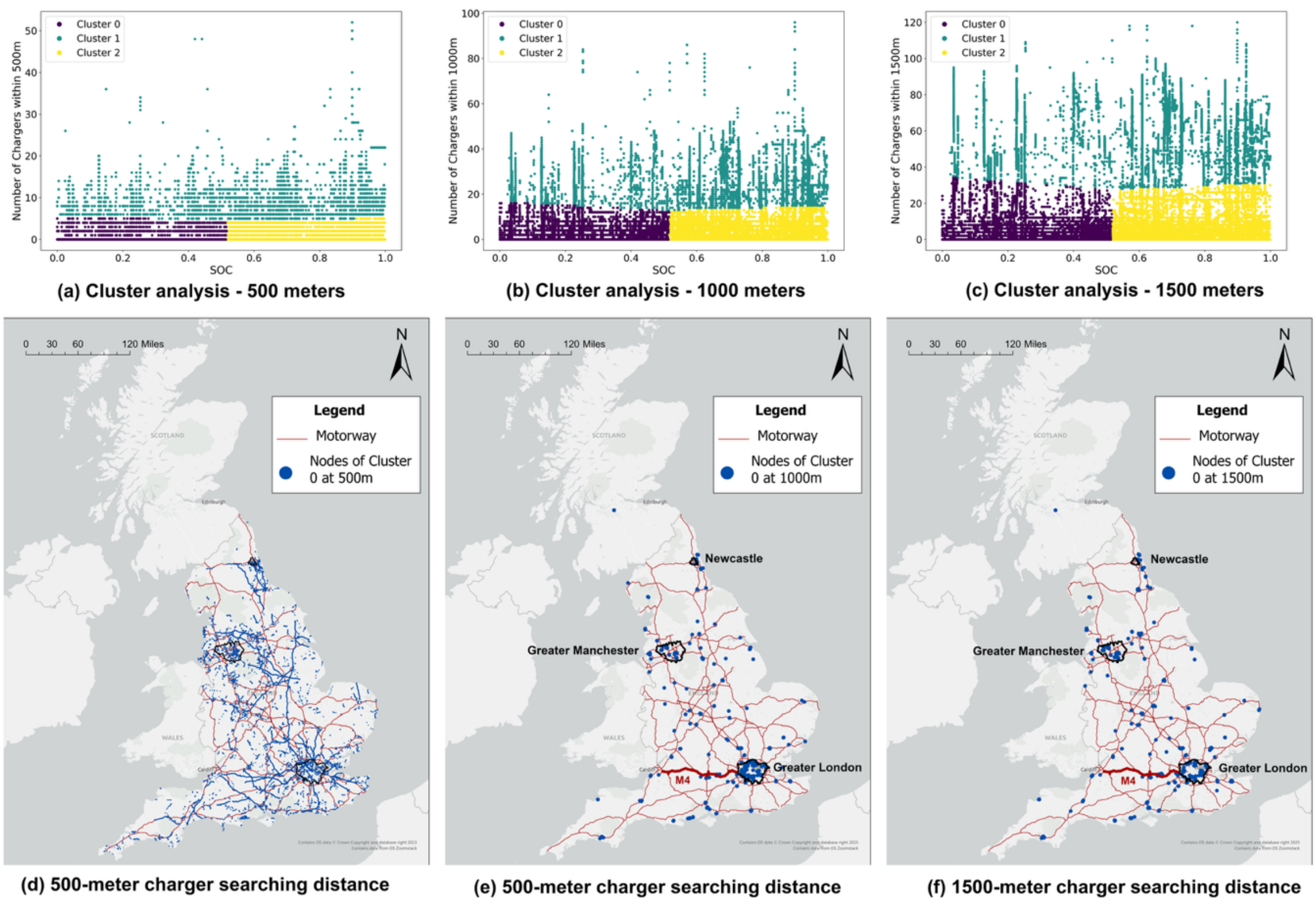

(a) Cluster analysis - 500 meters (b) Cluster analysis - 1000 meters (c) Cluster analysis - 1500 meters

(d) 500-meter charger searching distance (e) 500-meter charger searching distance (f) 1500-meter charger searching distance

Figure 3. Areas with high-risk of battery depletion

The 500-meter figure identified widely distributed high-risk areas. However, this distance likely overestimates infrastructure needs, as drivers typically accept detours longer than 500 meters for charging. In contrast, the 1000m and 1500m analyses revealed more realistic risk patterns. They both highlight areas such as Outer London, the eastern M4 corridor, Greater Manchester, and south Newcastle. The main difference between these two buffer distances is observed in Greater London. The 1500m result primarily highlighted high-risk areas along London's boundaries. The 1000m result identified more high-risk areas toward Inner London. Notably, compared to the SOC-only analysis shown in Figure 2(c), fewer areas along the M4 and M1 corridors were classified as high-risk. This suggests that, despite frequent low SOC observations, these corridors have relatively sufficient public chargers, which reduces the actual risk of battery depletion.

# 6 Discussion

This paper presents a large-scale, multi-stage agent-based RL model to train and simulate EV drivers' charging and driving behaviours in GB. Unlike previous studies that apply RL to optimise EV fleet operation strategies, we employ RL to capture the adaptive learning process of private EV drivers, who operate independently and have limited communication with each other. By validating each training episode against real-world charging data, we identify the simulation stage that best reflects reality, which also represents a balance between adaptive behaviour and bounded rationality. The validity of the model is further supported by the emergent behaviours of different types of trained agents, which align with established and intuitive understanding of EV driver behaviour. For example, leisure drivers show more flexible charging behaviours than commuters, while long-distance inter-city trips generate higher demand for en-route charging. These patterns demonstrate that the framework effectively captures realistic decision-making processes. Additionally, we identified the areas with higher risks of battery depletion based on the available EV charging station data and the simulation results.

The model also generated a high-resolution aggregated SOC distribution along road networks. It allows for the identification of locations with higher risks of battery depletion from both charging demand and infrastructure supply perspectives. Many areas with low average SOC are situated along inter-city motorways, likely due to the extensive distribution of long-distance travel in these areas. By exploring charging risks across multiple search distances, we identified key ‘charging deserts’ in GB. These places face both low SOC levels and limited charger provision, including boundaries of Outer London, Greater Manchester, south of Newcastle, and the eastern part of the M4 corridor. Charging failures in these areas could have greater risks for long-distance travellers. As a result, while initial charger deployment focused on destination and residential charging (Department of Transport, 2022), our findings support recent policy shifts toward developing rapid charging hubs along motorway corridors and urban peripheries in these areas (Department for Transport et al., 2022; Mohammed et al., 2024; Sirel and Brandt, 2023).

The identified charging demand distribution can also serve as a key input for infrastructure planning and decision-making models. As noted by Kchaou-Boujelben (2021), one of the major challenges in applying the public charger location models lies in accurately estimating the spatial and temporal distribution of charging demand. Recent studies have begun to address this limitation. For example, Yi et al. (2023) used rule-based ABMs to generate high-resolution driving profiles of EV drivers and predict where charging requests are likely to arise before optimising charger placement. The agent-based RL model proposed in this study can extend this direction of research by integrating RL-trained agents with adaptive behaviours and bounded rationality and providing more behaviourally realistic representation of charging demand distribution. These estimates can therefore be integrated into public charger optimisation frameworks, such as those by Anjos et al. (2020) and Ljósheim et al. (2026), to support more robust infrastructure planning decisions.

This paper also introduces a novel modelling framework that extends RL applications beyond fleet optimisation to understand how individual private EV drivers adapt their charging and driving behaviours. By simulating diverse charging patterns across a national-scale road network, the model goes beyond intra-city scenarios and captures the charging demand from long-distance, inter-city trips. Additionally, by validating the model with real-world charging session data, it effectively captures the learning process under bounded rationality and limited inter-communication between private EV drivers.

There are several limitations in this paper that worth further research. First, although the RL framework represents an avenue for modelling agent behaviours in unseen scenarios, questions remain regarding the trade-offs between model tractability. While we can explore EV drivers' final decisions in each episode, we cannot specify their exact reasons behind their choices. Second, due to computational resource limitations, this study is limited to simulating 1000 agents simultaneously. Future research could explore opportunities to apply the trained model to larger-scale applications with high performance computing infrastructure and simulate a greater number of EV drivers. Third, the NTS data includes England residents but lacks trips by Scotland and Wales residents. Future work could incorporate additional data to address this regional bias. The NTS dataset also does not provide information on EV ownership, and we assume that drivers maintain their existing travel patterns upon switching to EVs. Although previous studies have shown that using trip information from travel survey data in combination with appropriately specified EV attributes can effectively model EV charging behaviour (Pareschi et al., 2020), future research could integrate EV adoption inference models that account for socio-demographic factors, residential property types, and attitudes towards EVs to generate more representative synthetic EV driver populations. In addition, the initial SOC levels in this study are generated using a truncated normal distribution, which does not explicitly account for access to home and workplace charging. Access to home and workplace chargers are likely to be the most direct determinants of initial SOC. However, the NTS dataset does not include information on home or workplace charger ownership. It also does not include residential property types or workplace locations that could be used to infer such access. Future research could improve upon this approach by integrating data on home and workplace charger availability, once such data becomes available, to generate more realistic initial SOC distributions.

Furthermore, while the current framework captures heterogeneity in learning progress across agent types, future research could more explicitly model the mix of experienced and novice drivers within the EV driver population. This could be achieved by introducing rule-based agents with simple decision heuristics to represent novice drivers and combining them with RL-trained agents representing experienced drivers. Such approach would create a more explicitly heterogeneous agent population and allow for investigation of how the proportion of novice versus experienced drivers affects aggregated charging patterns. Additionally, public chargers can serve various user groups, including private EV drivers, EV fleets, and hired vehicles. Since this study only accounted for charging demand from private EV drivers, future studies could incorporate a broader range of user groups to provide a more comprehensive representation of public charger usage. Moreover, the model assumes that drivers follow the shortest path to their destinations as determined by Dijkstra's algorithm. While this ensures computational efficiency, real-world drivers may deviate from optimal routes due to personal preferences, habitual behaviour, or incomplete information. Future research should incorporate bounded rationality in route choices through approaches such as stochastic shortest path formulations. This enables a more realistic reflection of driving and route choice behaviour of EV drivers. Finally, the charger location data used in this study is limited in coverage and may not include all public chargers across GB. This limitation may affect the accuracy of identifying "charging deserts" and the overall representation of charging accessibility. Once a more complete, nationwide dataset on public charger data becomes available, it would allow for a more detailed and reliable exploration of charging deserts.

# 7 Conclusion

To address the limitations of data-driven or simulation-based approaches, we developed a agent-based RL model using DQN to capture the diverse behaviours of EV drivers in GB. The modelling framework captures both the adaptive charging behaviours of EV drivers and the emergent patterns that arise from interactions at charging stations. Rather than seeking optimal charging strategies, which rarely happens for private EV drivers, we validated the model against real-world charging session data, and identified a group of episodes that closely align with observed charging behaviours. The validity of the model is further supported by the emergent behaviours of trained agents, which align with common-sense expectations. Additionally, our results provide a validated modelling framework for estimating spatial distribution of charging demand and identifying areas at high risk of battery depletion. It also highlights the importance of establishing more fast charging hubs along motorway corridors in the next phase of public charger deployment.

# Supplementary Materials

## Data and study area

### 1 Data for simulation environment

We use the road network data from the OS Open Roads dataset (Ordnance Survey, 2023b), and public charger data from the ChargePoint dataset (ChargePoint, 2023) to establish the modelling environment. The public charger data is scraped from ChargePoint, an EV charging station provider operating in North America and Europe. It includes information from multiple charger providers across GB. While using web scraping to gather the details of public chargers, we ensured that our activities were limited to publicly accessible information and conducted solely for non-commercial research purposes. Our scraping process strictly adhered to the web scraping policy by the Office for National Statistics (Office for National Statistics, 2020) and complied with UK copyright law, which permits lawful access to material for non-commercial research purposes (Intellectual Property Office, 2021). The attributes of the chargers are listed in Table S1, and the distribution of all chargers are shown in Figure S1. The dataset includes 17,114 chargers with a total of 25,923 ports across GB, with each charger contains one or more ports.

Table S1. Summary statistics for public charger attributes.

| Name | Definition | Value range |
|---|---|---|
| Charger ID | Unique identifier for each charger. | - |
| Port number | The number of charging ports at each charger. | 1-12 |
| Location | The geographic coordinates (latitude and longitude) of the charger. | Latitude: 50.1-59.0; Longitude: -7.2-1.7 |
| Price (£/KWh) | The cost of electricity per kilowatt-hour for charging. | 0.3-1.1 |
| Speed (KW) | The charging speed of the charger in kilowatts. | 6.4-300.2 |
| Initial parking fee (£) | The parking fee for the first hour during charging. | 0.0-0.6 |
| Additional parking fee (£/h) | The parking fee for each additional hour during the charging session. | 1.2-24.0 |

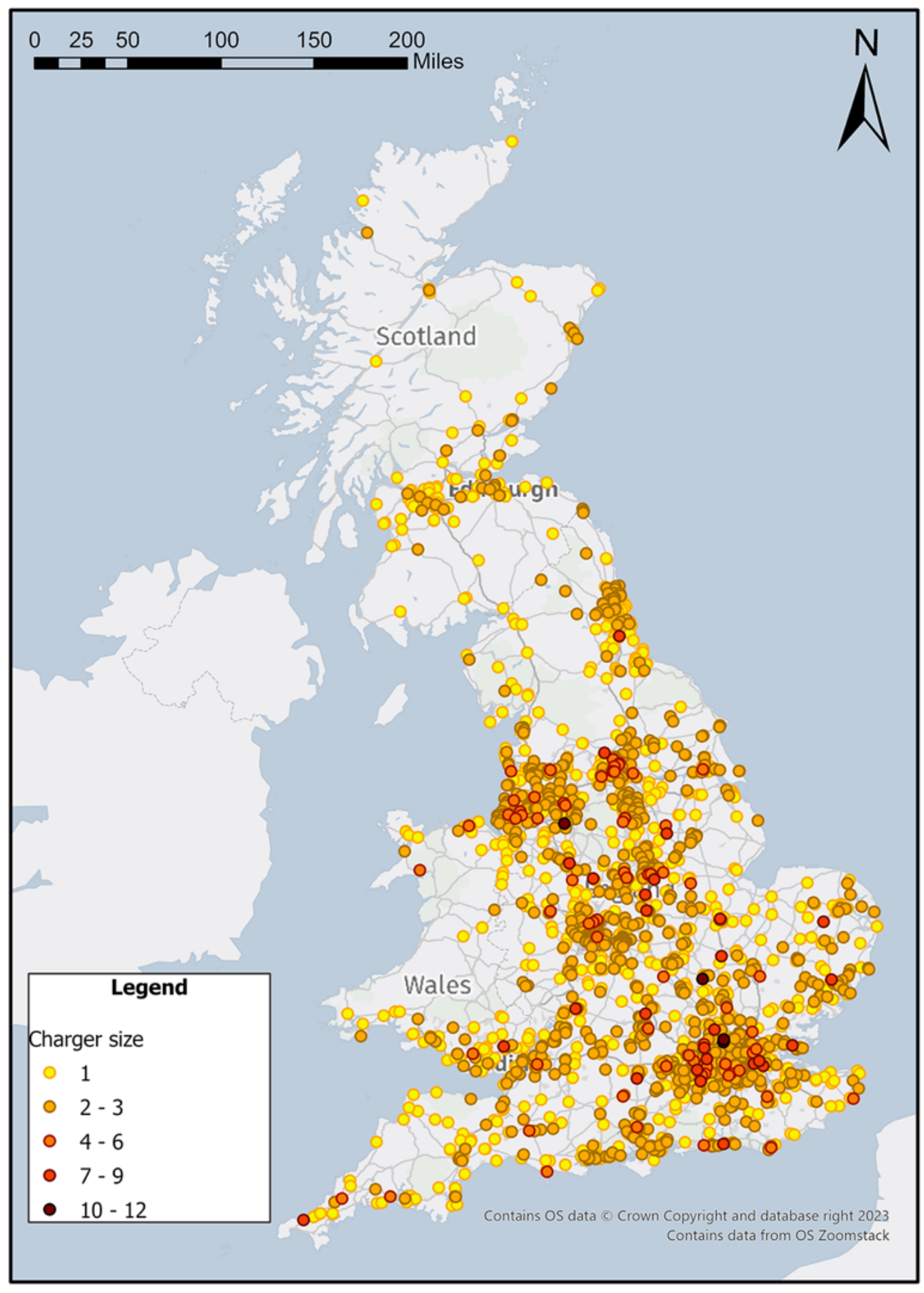

Figure S1 Distributions of public chargers in GB.

## 2 Data for EV driver agents

This section describes the data sources and methods used to construct EV driver population for simulation. Pareschi et al. (2020) have found that travel diaries drawn from travel survey data, together with appropriate settings of EV-related attributes, can provide a reliable and accurate foundation for simulating EV driver behaviour and charging loads. Based on this foundation, we formulate synthetic EV driver populations based on trip schedules from NTS dataset, and EV-specific attributes generated based on recognised range and distributions from previous studies and EV type datasets. Data sources of different variables are summarised in Table S2 and explained in detail in this section.

Table S2 Data sources to formulate EV population

| **Variable Name** | **Description** | **Processing details** |
| --- | --- | --- |
| | **Trip schedule from NTS dataset** | |
| Trip purpose | The purposes of the trip | Specific reclassification methods are included in the Data Pre-processing section. |

| Origin unitary authority | The unitary authority the trip origin belongs to | These three variables (origin, destination, and trip distance) are combined with the OS Code-Point data to determine the specific origin and destination coordinates of the trip, as well as the original shortest paths without charging activities for simulation. Details are included in the Data Pre-processing section. |
|---|---|---|
| Destination unitary authority | The unitary authority the trip destination belongs to | |
| Trip distance | Total trip distance by vehicles excluding short distance walks (miles) | |
| Trip start time | Time in minutes past midnight when the trip starts | Directly derived from the dataset. |
| **EV-specific attributes** | | |
| Initial SOC | The SOC level when driver start the trip over a day. | Generated based on truncated normal distribution bounded within the interval (0,1) (Pareschi et al., 2020; Yang et al., 2025). |
| Battery size and Battery consumption rate | Extracted from the EV Database website (EV Database, 2024) | Each EV driver is randomly assigned a vehicle type available on UK's market when initialising the model. |
| Charger density surrounding each driver | Number of public chargers within 500m buffer of shortest path, normalised by trip length. | Calculated from charger locations and route geometry. |
| Trip co-occurrence density surrounding each driver | Number of concurrent trips intersecting the route's 500m buffer, normalised by trip length. | Calculated from trip schedules and route geometry. |

## Trip Schedule

Since EV-specific travel data is rarely available due to privacy concerns and limited market penetration, we follow the common approach in EV simulation studies of using general travel survey data to represent EV driver travel diaries (Pareschi et al., 2020; Yan et al., 2020). It should be noted that NTS dataset does not identify EV owners; rather, we use the travel patterns from the survey and assume that drivers maintain their existing travel plans upon switching to EVs (Yan et al., 2020). Trip schedule information derived from NTS dataset includes trip purposes, origin and destination unitary authorities, trip distance, and trip start time.

**Initial SOC**

Drivers may start the day with different initial SOC levels. These differences can be influenced by factors such as home charger availability (Liao et al., 2023) and access to workplace charger (Andrenacci and Valentini, 2023). Drivers with access to either a home or workplace charger are more likely to start their trips with a higher State of Charge (SOC) than those without access. However, home and workplace charging is not considered in this research due to the lack of data on home and workplace charger accessibility in the UK. Obtaining such data is challenging due to privacy concerns and the fragmented nature of data across multiple charger providers. One potential approach to inferring home charger availability is through residential property type. For example, residents of houses are more likely to have the option to install home chargers, whereas those living in flats often do not. (Visaria et al., 2022; Xu et al., 2017; Chakraborty et al., 2019). Nevertheless, the NTS dataset does not include information either on home charger ownership or property types.

Given these data constraints and following established practice in EV simulation literature (Pareschi et al., 2020; Yang et al., 2025), we initialise each EV with a random initial SOC level based on a truncated normal distribution bounded within the interval (0,1). This approach captures population-level heterogeneity in initial SOC while avoiding assumptions about the relationship between individual socio-demographic features and charging behaviour, which is hard to be validated with available data.

**Battery size and battery consumption rate**

The EV attribute data is extracted from the EV Database website (EV Database, 2024), which provides information on vehicle make and model, battery range, and consumption rates of EVs available in the UK market. As of February 18, 2024, there are 231 EV types included in the dataset. Each EV driver is randomly assigned a vehicle type when initialising the model. Since the primary objective of this study is to analyse driver behaviour instead of technical aspects of vehicle performance, we do not account for the effects of battery aging or varying consumption over time. Instead, the specifications used in the simulation reflect those of newly manufactured EVs. Table S3 includes some examples of these EV models in the dataset.

Table S3. Examples of EV models in the UK.

| EV Name | Battery Size (kWh) | Consumption Rate (Wh/mi) |
|---|---|---|
| **Tesla Model Y** | 57.5 | 267.0 |
| **Tesla Model 3** | 57.5 | 221.0 |
| **Kia Niro EV** | 64.8 | 270.0 |
| **Volkswagen ID.3** | 59.0 | 268.0 |
| **Nissan Leaf** | 39.0 | 269.0 |

## Environmental factors

Besides individual preferences and trip-specific features, a driver's behaviour can also be influenced by its environmental factors. These include the distribution of the public chargers nearby and the presence of other EVs traveling within overlapping time windows. Therefore, after identifying the origin and destination (OD) coordinates and calculating the shortest paths between them (as detailed in Table S2 and the Data Pre-processing section), we introduce two environmental indicators to quantify charging infrastructure availability and potential competition for public chargers among drivers.

Charger density (CD) is defined as the number of public chargers within a 500-meter buffer along the shortest path, normalised by the total path length. This metric represents the relative availability of charging infrastructure along a driver's route. As shown in Equation (1), $d_i$ is the distance of from charger $i$ to the shortest path. $L_{trip}$ is the length of the shortest path. $N_c$ is the total number of public chargers considered, and the indicator function $1(d_i \leq 500)$ equals 1 if charger $i$ lies within 500 meters of the shortest path, and 0 otherwise.

$$CD = \frac{\sum_{i=1}^{N_c} 1\ (d_i \leq 500)}{L_{trip}} \tag{1}$$

Trip co-occurrence density (TCD) is defined as the number of concurrent trips intersecting the route's 500-meter buffer zone during the travel time window, normalised by trip distance. It quantifies the potential charging competition along the trip. As denoted in Equation (2), $N_t$ is the total number of concurrent trips within the travel time window. $R_j$ is the route of the concurrent trip from other drivers. $B$ is the 500-meter buffer zone around the shortest path.

$$TCD = \frac{\sum_{j=1}^{N_t} 1\ (R_j \cap B\ \neq\ \emptyset)}{L_{trip}} \tag{2}$$

### 3 Data for model validation

We use charging session data from ChargePoint to validate the model. Besides the attributes detailed in Table S1, we also scraped the availability status of chargers, which includes statuses of 'In-use', 'Available', and 'Closed'. The status data is collected at 15-minute intervals from December 5, 2023, to May 15, 2024. Figure S2 shows the total charger usage time change in GB for a sample week in February. On average, peak usage hours occur between 12:00 and 14:00.

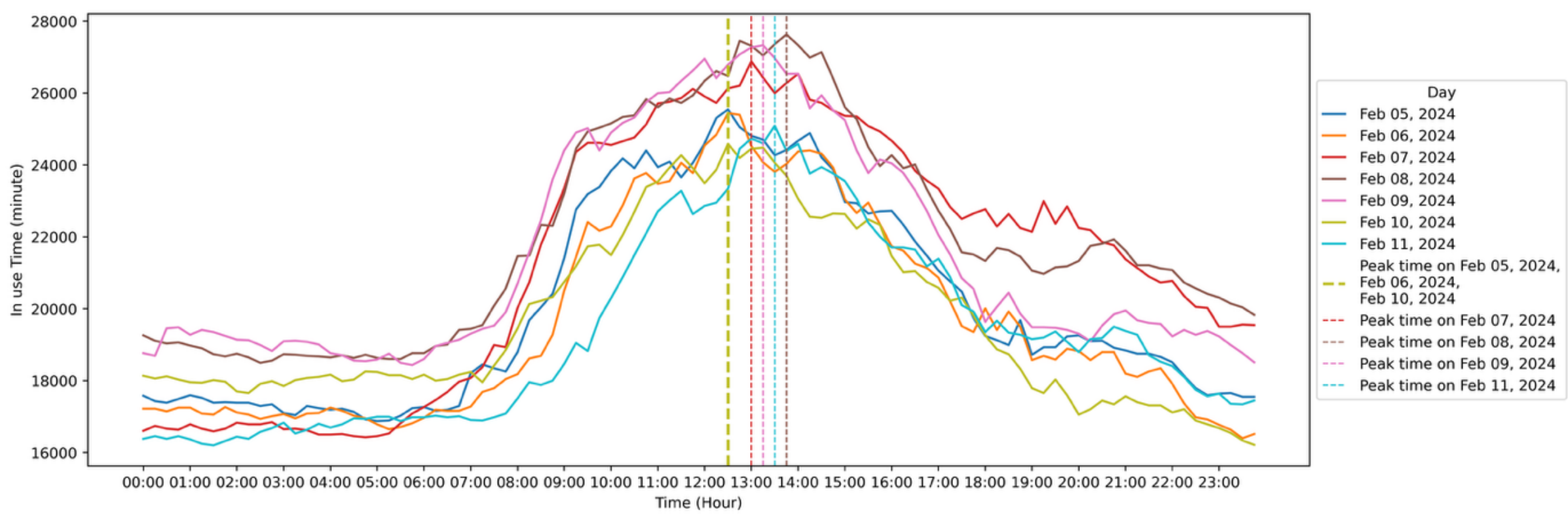

Figure S2. Total usage time of all public charger in GB in a sample week (Feb 5-11, 2024).

## Data Pre-processing

We reclassified the trip purpose variable in the original dataset into two categories: work and leisure. Specific reclassification methods are shown in Table S4.

Table S4 Reclassification of trip purposes

| **Original trip purposes** | **Reclassified trip purposes** |
|---|---|
| Commuting | |
| Personal business | Work |
| Business | |
| Education/escort education | |
| Other escort | |
| Shopping | Leisure |
| Leisure | |

We assume that the vehicle drivers in the NTS data have adopted EVs while their trip OD remain unchanged. Therefore, the variables of trip origin unitary authority and trip destination unitary authority from the dataset are selected to formulate the OD information of the trips of EV drivers. As these OD information at the unitarity authority level is not specific enough to serve as the OD of EV drivers when initialising the model, OS Code-Point data (Ordnance Survey, 2023a), which locates over 1.7 million postcode units for GB and are unique references for identifying addresses, are combined with the NTS data to determine the start and end coordinates for each trip in the model. The process is designed to ensure that the OD distance in the model closely match the trip distance values in the dataset, while also reducing the computational burden associated with calculating large OD matrices.

The selection of origins and destinations for each trip was carried out based on the UK postcode system. The UK postcode system follows a hierarchical structure (Table S5). A full postcode consists of two parts: the outward code and the inward code. The postcode area is the initial letter(s) to represent a major town or city. The postcode district adds one or two numbers or letters to the area, typically covering part of a town or urban area. The postcode sector includes the district plus the first digit of the inward code, usually representing a few thousand addresses. Finally, the postcode unit is the full postcode, identifying approximately 15–20 addresses on average.

Table S5. UK postcode structure.

| Postcode | | | |
|---|---|---|---|
| Outward code | | Inward code | |
| Area | District | Sector | Unit |
| Example (SW1A 2AA) | | | |
| SW | SW1A | SW1A 2 | SW1A 2AA |

As indicated in Figure S3, the starting point of each trip $P_{orig}$ is randomly selected from the postcodes within the trip start unitary authority. Using Dijkstra's shortest path algorithm, the distances from $P_{orig}$ to centroid of each postcode district are calculated. The district that minimises the absolute difference with the trip distance value $T$ from the dataset is chosen as the target postcode district. Then, similar to the previous step, the destination postcode sector is determined. Finally, among all the postcode points within the target postcode sector, $P_{dest}$ is determined as the postcode point that has the minimal distance difference with $T$.

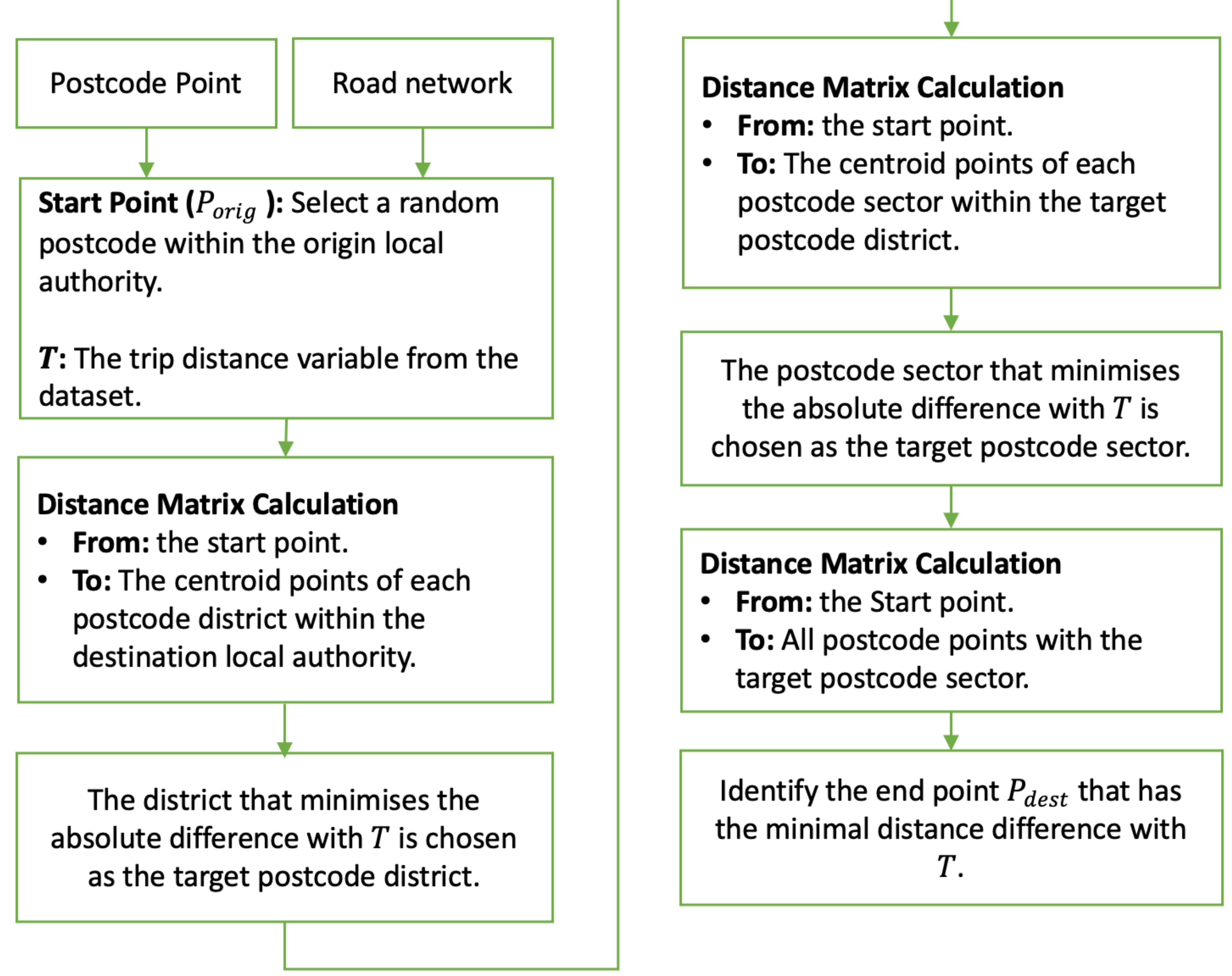

Figure S3. Flowchart of selecting origin and destination for each trip.

## Justification of Reward Function Design

The reward function (Equation 13 in the main text) was designed to capture key trade-offs faced by real-world private EV drivers when making charging decisions. It collectively considers four components: SOC threshold impact ($D_{SOC}$), charging urgency factor ($D_{charge}$), cost impact ($D_{cost}$), and trips completion status ($Status$). The design of the reward function is based on established understanding of EV driver decision-making from the literature and practical considerations for realistic behaviour simulation. The rationale for each component is specified below.

**Trips completion status ($Status$)**

The primary objective of an EV driver is to complete their planned trips. The $Status$ variable ensures that this fundamental goal is explicitly valued in the reward function. Successful trip completion provides an additional reward, while running out of charge before reaching the destination incurs a penalty. Without this component, agents might learn strategies that minimise charging costs at the expense of failing to complete trips.

This formulation reflects real-world behaviour, that drivers will not adopt strategies that risk being stranded due to battery depletion. Previous studies show that EV drivers show forward-looking properties and planning abilities when scheduling charging and driving (Thorhauge et al., 2024). Additionally, compared to other factors, EV drivers perceive completing the planned trips the primary factor influencing their charging decision (Ge and MacKenzie, 2022).

**SOC Threshold Impact ($D_{SOC}$)**

Range anxiety is identified as a critical determinant of EV drivers' charging behaviour (Chaudhari et al., 2019; Li et al., 2025; Ling et al., 2022; van der Kam et al., 2019). Building on the concept of 'safety buffer' (Kurani et al., 1994), Franke et al. (2012) introduced the concept of 'comfortable range', which is defined as the subjective portion of battery capacity a driver is willing to use without experiencing stress. Rainieri et al. (2023) further show that the greater the discrepancy between available battery charge and a driver's preferred safety battery charge level, the more likely drivers are to experience discomfort and stress.

To capture this behavioural pattern, $D_{SOC}$ is defined as the difference between the current SOC and the driver's psychological threshold ($SOC_{threshold}$). A larger positive difference indicates greater perceived comfort and results in a higher reward, whereas a negative difference reflects increased anxiety and reduces the reward accordingly.

**Cost Impact ($D_{cost}$)**

EV drivers typically consider both monetary and time costs when making charging decisions (Wang et al., 2024; Yang et al., 2020), although the relative importance of these costs varies by drivers' preferences and travel contexts (Daina et al., 2017). To reflect this heterogeneity, each agent is assigned with individual sensitivities to financial ($\alpha$) and time costs ($\beta$). In addition, the exponent $n/m$ is introduced to account for the interaction between charging frequency, trip length, and perceived disruption. Specifically, frequent charging imposes a greater perceived burden during short trips than during long-distance journeys, where charging stops are more expected and accepted. This design enables the reward function to differentiate cost impacts across travel contexts.

**Charging Urgency Factor ($D_{charge}$)**

Charging urgency reflects drivers' perceptions of risk and opportunity associated with charging timing and infrastructure availability. Tate et al. (2008) indicate that EV drivers have pervasive concerns about becoming stranded with a depleted battery, particularly when operating away from charging infrastructure. Additionally, unlike refuelling conventional vehicles, EV charging requires extended dwell times at charging stations. Helmus and Wolbertus (2023) indicate that one user's access to charging station reduces availability for others. In stochastic environments where private EV drivers have limited information about others' intentions, drivers can encounter congestion and queuing at certain chargers. To capture this effect, the reward function incorporates $chance_{charge}$, defined as the ratio of charger density (CD) to trip co-occurrence density (TCD). This ratio represents the perceived likelihood of encountering an available charger relative to potential competition. Higher values indicate a greater perceived chance of successful charging without queueing.

Additionally, the timing dimension of charging urgency depends on both the battery level at trip start ($status_{battery}$) and the timing of charging relative to the trip start ($timing_{charge}$). Lower initial battery levels increase perceived urgency, as drivers face a higher risk of depletion and reduced flexibility in charging choices. In contrast, $timing_{charge}$ , which captures when charging occurs relative to trip start time, and the battery level when trips start ($status_{battery}$). Lower values correspond to early charging, which reduces perceived urgency as drivers secure charging opportunities proactively. This behaviour aligns with findings by Thorhauge et al. (2024) that with forward-looking planning tendencies, many EV drives charge at early stages of trips even when there is sufficient battery charge remains.

## Clustering Results

The cluster analysis was conducted based on four continuous features: total trip distance, initial battery level, trip co-occurrence density, and public charger density. Five clusters were identified based on these features. To capture behavioural differences associated with trip purpose, each of the five clusters was then subdivided by work and leisure labels. Instead of incorporating categorical variables directly into the initial clustering process, this approach ensures more accurate distance-based calculations while still capturing the behavioural differences between work and leisure trips. The centroid of the resulting clusters is listed in Table S6.

Table S6 Classifications and centroids of clustering analysis

| Cluster | Trip purpose | Public charger density | Trip co-occurrence density | Total trip distance | Initial battery level | Cluster feature |
|---|---|---|---|---|---|---|
| 1 | work<br>leisure | 0.03 | 0.03 | 19.59 | 0.49 | Mid-distance inter-city drivers |
| 2 | work<br>leisure | 0.03 | 0.04 | 30.07 | 0.25 | Long-distance Low-battery inter-city drivers |
| 3 | work<br>leisure | 0.26 | 0.31 | 20.42 | 0.52 | Mid-distance drivers in high-density areas |
| 4 | work<br>leisure | 0.03 | 0.04 | 30.62 | 0.76 | Long-distance inter-city drivers |
| 5 | work<br>leisure | 1.00 | 0.49 | 10.61 | 0.70 | Short-distance drivers in high-density areas |

# Model Validation

The model was validated using 21 days of real-world charging session data. Cross-correlation analysis was performed between the simulated results of each episode and the daily charging session data, both spatially and temporally. The correlation results are shown in Figure S4. In the figures, the shaded area represents the range of correlation values observed across different model runs with different sets of training and simulation agents, and the solid line indicates the mean correlation. This variation in correlation arises from the use of different combinations of training and simulation agents across model runs to ensure the robustness of the results.

We first perform a Pearson correlation analysis between spatial distribution of charger usage for each model runs and each of the 21 days of real-world charger usage data. The spatial correlation value stabilised after around the $350^{th}$ episode, with average values ranging between 0.74 and 0.75 in later episodes.

Model validation is also conducted on a temporal scale. The temporal cross-correlation result shows a more dynamic pattern compared to the spatial validation, with a significant upward trend during the early episodes, followed by a stable phase between the $350^{th}$ and $750^{th}$ episodes, and a gradual decline thereafter. Episodes between 350 and 750 marked the peak phase of temporal correlation, with values fall between 0.65 and 0.70. This indicates strong model performance. Compared to ABM with static behaviour rules by Feng et al. (2024), which worked on similar modelling purposes for private EV driver behaviours and used the same validation method, our higher values highlight the advantage of RL in replicating real-world scenarios.

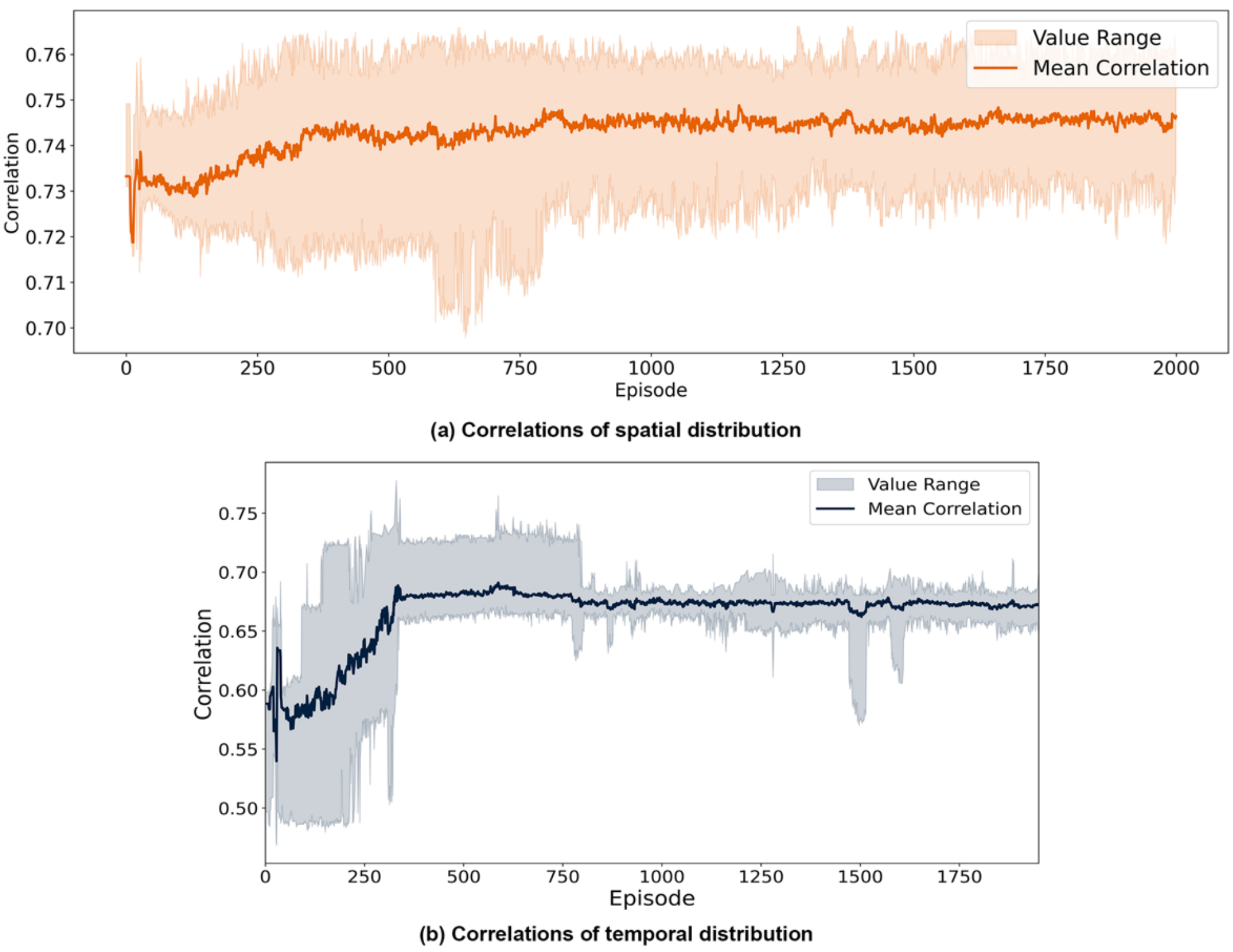

(a) Correlations of spatial distribution

(b) Correlations of temporal distribution

Figure S4. Correlation results

# Model Calibration

To ensure the model's effectiveness, we calibrated three key parameters: learning rate, discount factor, and exploration rate. Each parameter was calibrated independently, with the calibrated values presented in Table S7. The impact of these parameters on reward changes across all vehicle types is illustrated in Figures S5-S7, respectively.

The learning rate significantly affects the model's learning stability and convergence speed. As shown in Figure S4, a learning rate of 0.0001 resulted in greater stability, with fewer fluctuations in the reward change and a smoother learning process.

Table S7. Parameters and Values.

| Parameter | Values |
| --- | --- |
| Learning rate | [0.001, 0.00075, 0.0005, 0.00025, 0.0001] |
| Discount factor | [0.9, 0.95, 0.99], |
| Exploration rate range | [0.99→0.01, 0.98→0.01, 0.97→0.01, 0.96→0.01, 0.95→0.01] |

We also calibrated the discount factor which determines the importance agents place on future rewards. Higher discount factor can make driver values more to future rewards. A higher discount factor encourages agents to prioritise long-term rewards. Among the candidate values, 0.95 was selected as it enables agents to effectively balance immediate cost savings and future charging needs.

Lastly, the exploration rate was calibrated with different starting values ( $\epsilon_{max}$ = 0.99/0.98/0.97/0.96/0.95). Following an $\epsilon$-greedy strategy, a high initial exploration rate was chosen to allow agents to explore the environment more thoroughly by selecting random actions. To ensure agents exploit their learned knowledge, all $\epsilon$ decays over the 2000 training episodes and reaches 0.01 at the end of training.

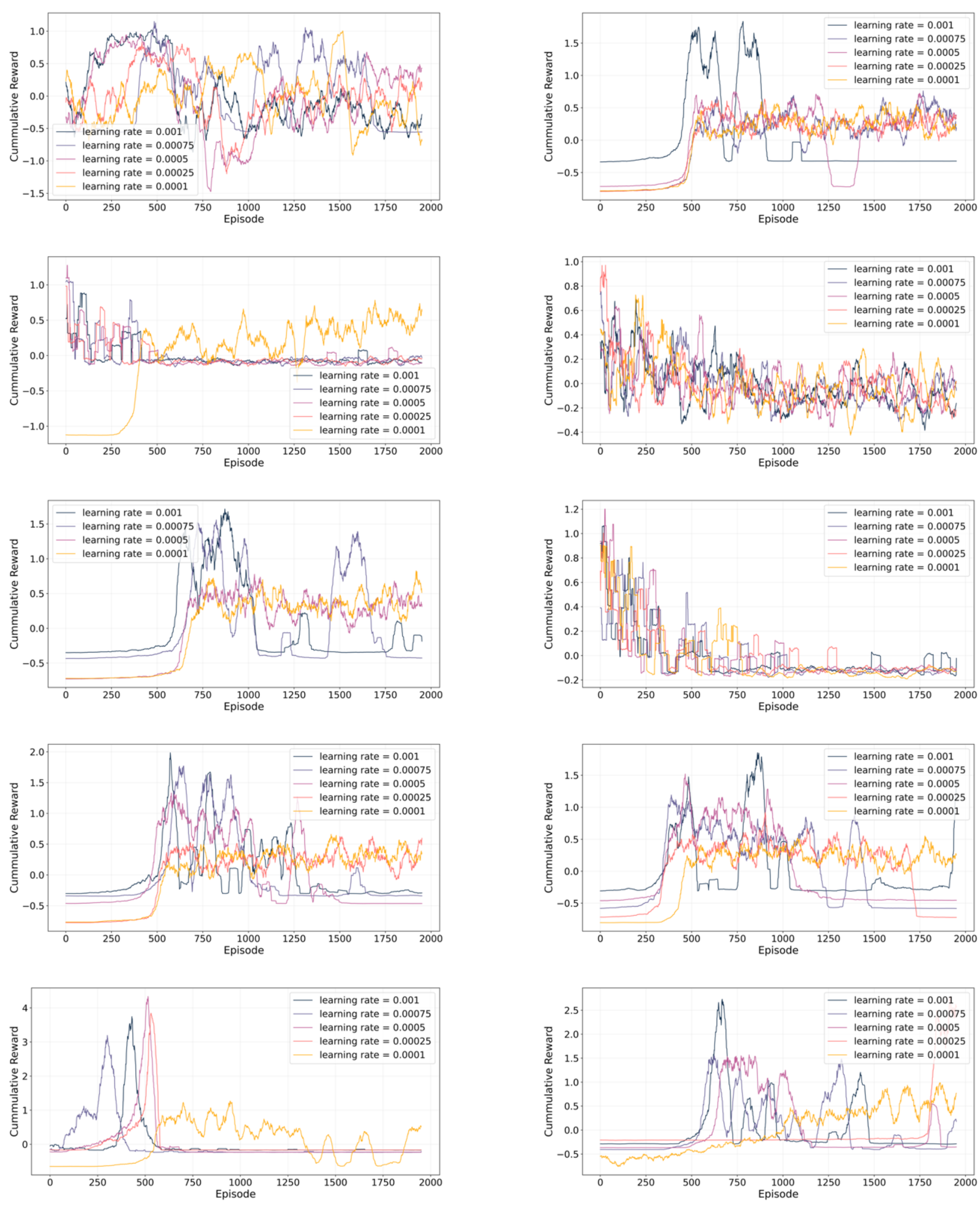

Figure S5 Parameter calibration of learning rate.

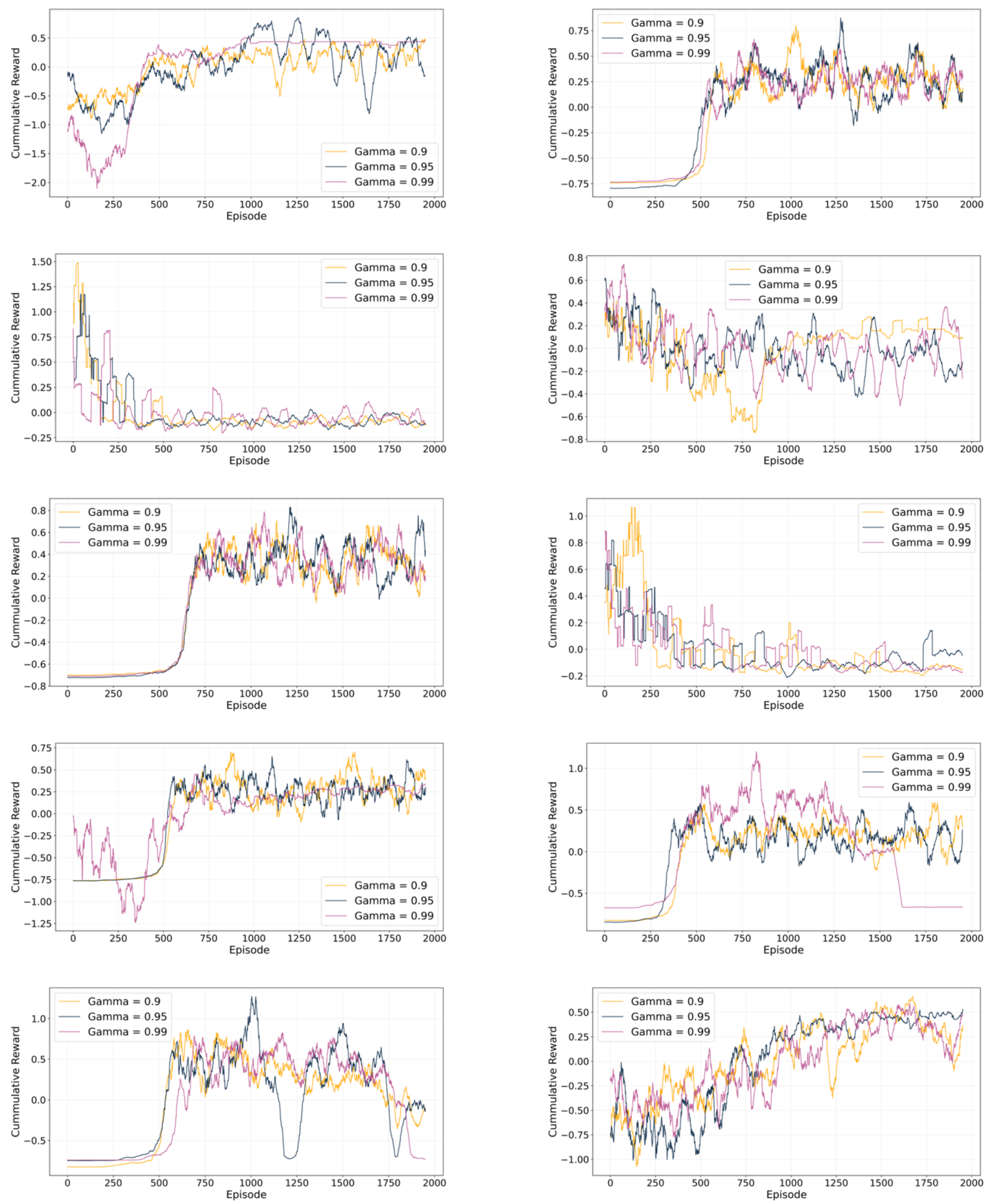

Figure S6 Parameter calibration of discount factor.

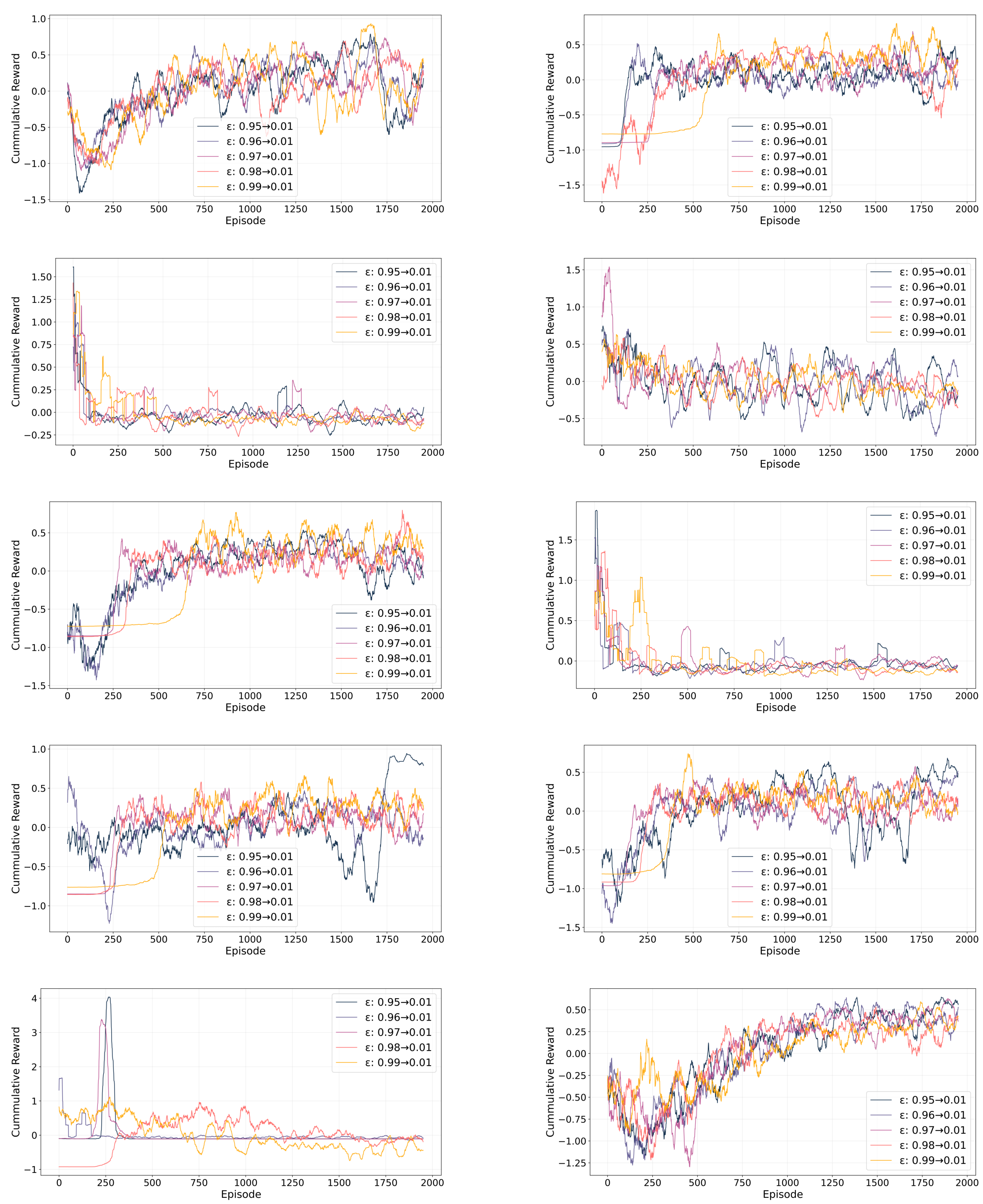

Figure S7 Parameter calibration of exploration rate.

# Cumulative Reward Change

The cumulative reward change for each type of training agent across all episodes is shown in Figure S8. The shaded range in the figure represents the variation in reward outcomes across different combinations of training and simulation agents. For agents in Cluster 1 (work and leisure), Cluster 3 (work), Cluster 4 (work and leisure), and Cluster 5 (leisure), an upward trend in cumulative rewards suggests that they progressively learned more optimal charging strategies over time. However, for agents in Cluster 2 (work and leisure) and Cluster 3 (leisure), a decrease in cumulative rewards upon convergence indicates that competition for chargers forced them to adopt less optimal strategies.

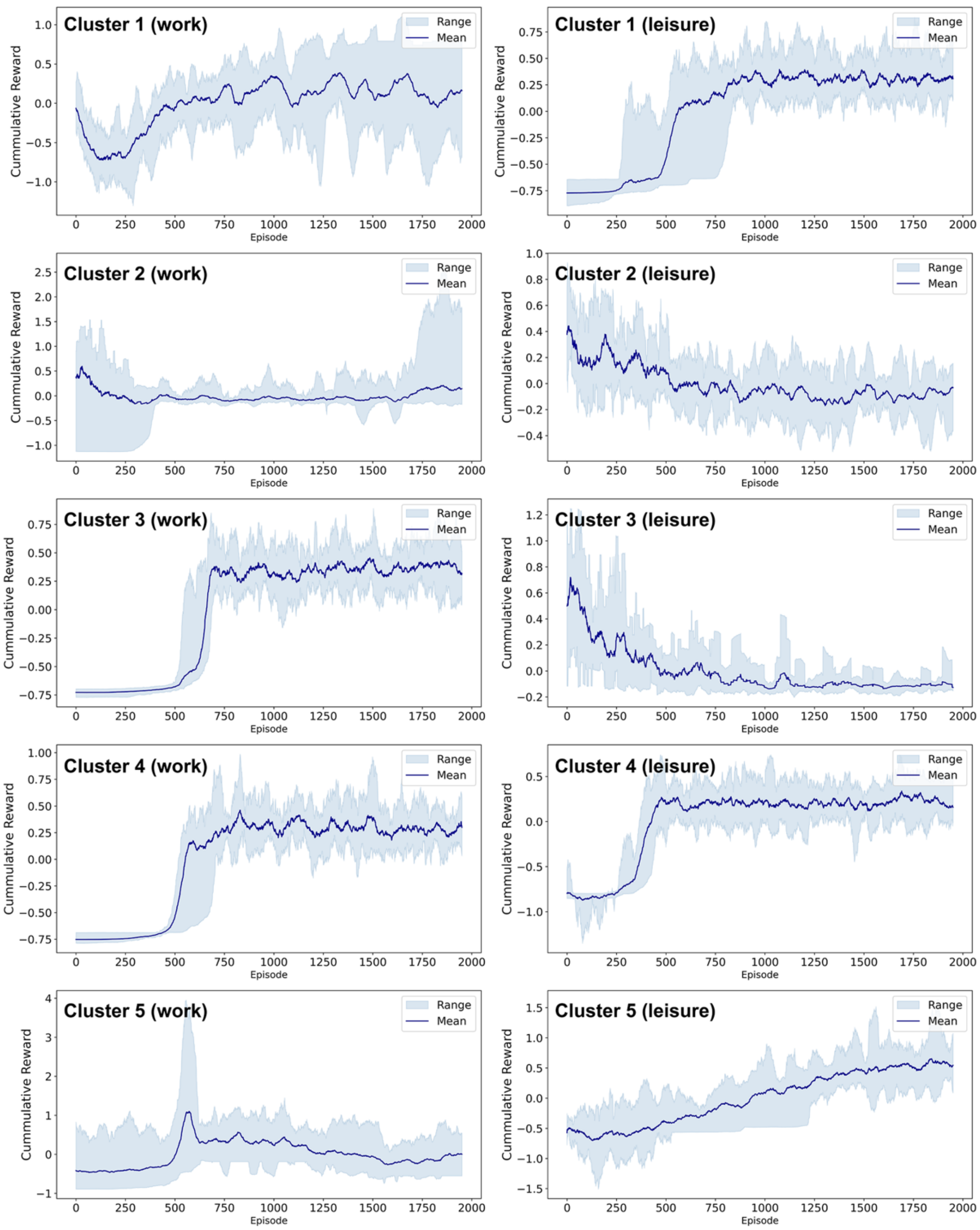

Figure S8. Reward change for ten cluster agents.

# Action selection Possibility of Agents

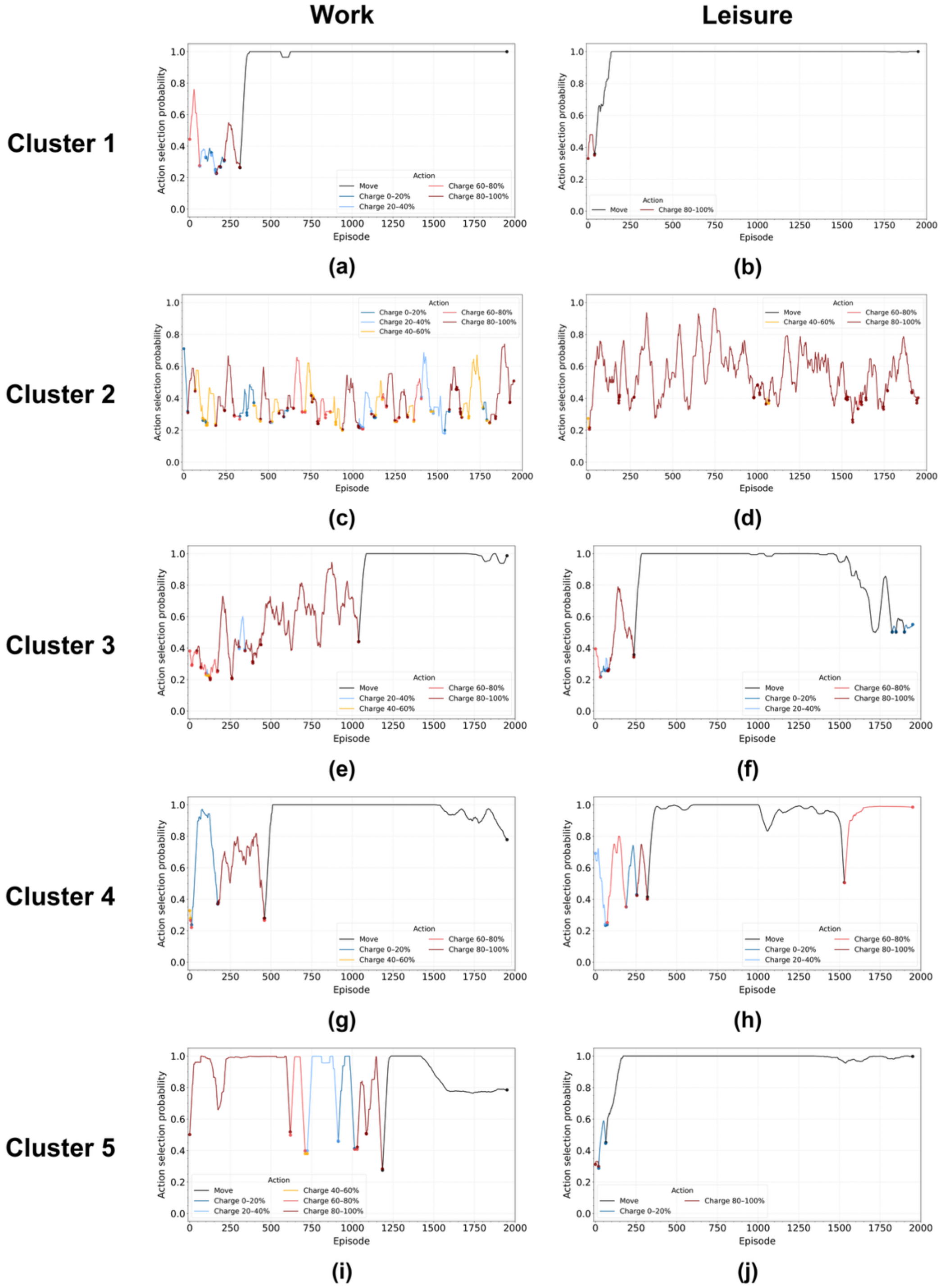

Figure S9. Dominant action selection for Group 1 drivers in Cluster 1 - 5.

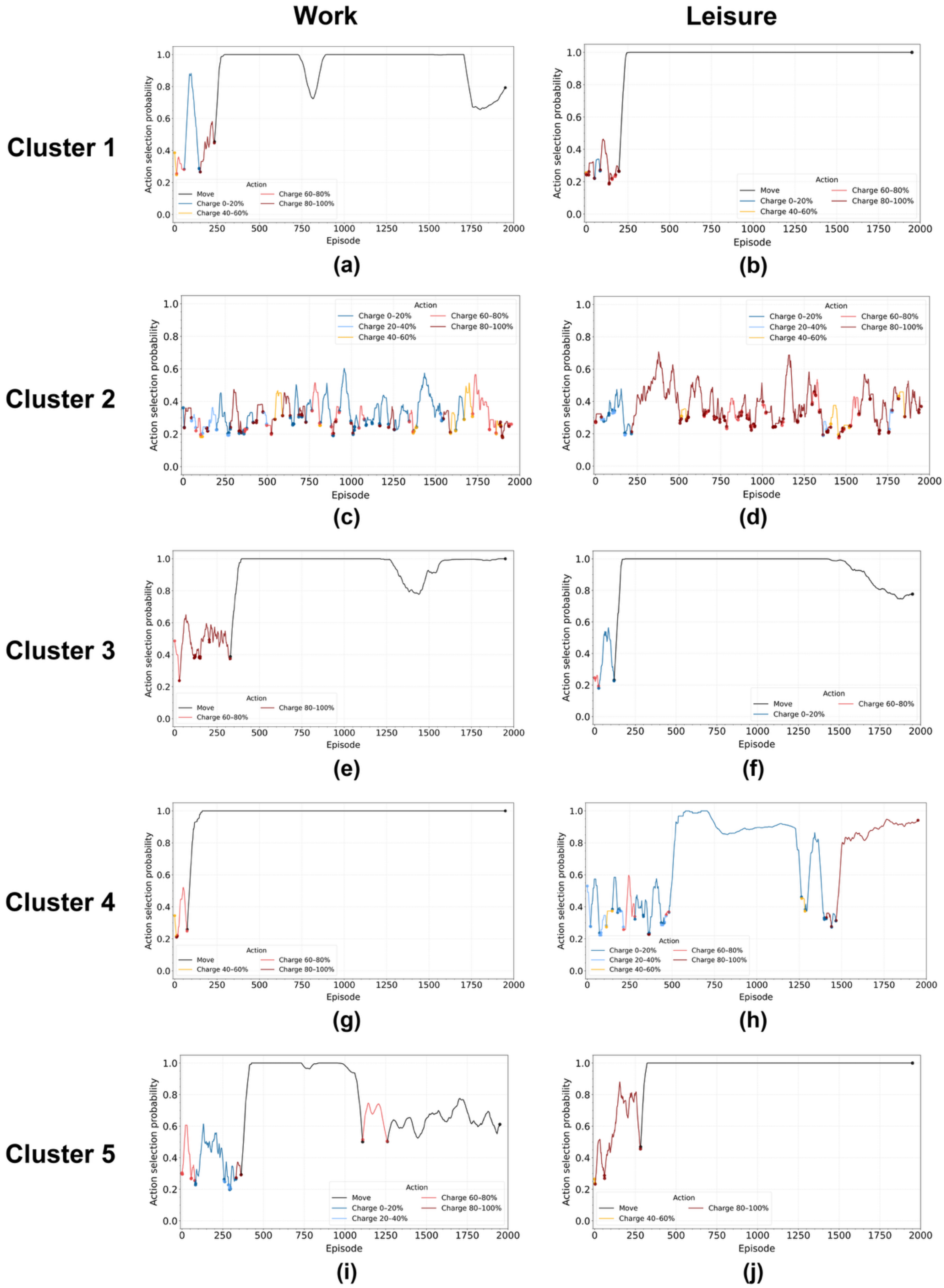

Figure S10. Dominant action selection for Group 2 drivers in Cluster 1 - 5.

There are two distinct groups of agents used for training. Figure S9 show the actions with the most frequently selected actions for one group of agents over 2000 episodes. The action selection possibility of another group of training agent is shown in Figure S10.

Figure S9 (a) and (b) shows the training agents from Cluster 1. These EV drivers start with a medium battery level and undertake medium-distance trips (see Table 1). They face low competition levels due to low TCD and limited charger availability in their surroundings. Regardless of travel purposes, their action choice stabilises after approximately 250-300 episodes when they tend to avoid charging.

The action choices for drivers in Cluster 2 are shown in Figure S9 (c) and (d). These drivers start with low battery levels, undertake long-distance trips, experience low competition but face limited charging opportunities. Over the course of training, work drivers progressively adjust their charging behaviours, ultimately choosing high charge amounts (70%–100%). In contrast, leisure drivers show more consistent behaviours throughout training and end with slightly higher charge levels (90%–100%), which reflects their greater schedule flexibility for longer time of charging. This scenario highlights the necessity of charging for long-distance travel. Drivers must recharge to avoid running out of power and to mitigate range anxiety, which is a common concern for long-distance journeys.

The action choices of drivers in Cluster 3 are shown in Figure S9 (e) and (f). These drivers start with medium battery levels, undertake medium-distance trips, and encounter significant competition but have sufficient access to charging infrastructure. These agents may be found in major urban areas, where there are high density of chargers and large number of EV drivers. In latter training episodes, work commuters tend to avoid charging due to potential long wait times at stations. In contrast, leisure drivers either proceed directly to their destinations or charge moderately, as their flexible schedules allow them to manage range anxiety more effectively.

The action choices of agents in Cluster 4 are shown in Figure S9 (g) and (h). These drivers start with high battery levels, encounter fewer surrounding EVs, and have limited charging opportunities. A notable difference is observed between work commuters and leisure travellers. Work commuters travel directly to their destinations without stopping to charge. Leisure drivers make charging stops, even when their battery level is sufficient to complete their trips. These results suggest that when drivers have enough battery to complete their trips, those on tight schedules—such as work commuters—prioritise reaching their destinations without delay. In contrast, leisure drivers, who have more flexible schedules, tend to stop for charging, potentially as a way to reduce range anxiety, even if it is not strictly necessary.

The action choices of agents in Cluster 5 are shown in Figure S9 (i) and (j). These drivers also start with high battery levels but take short-distance trips. Unlike Cluster 3, they face higher congestion levels and have the most extensive access to charging infrastructure. These drivers are likely to be located in city centres. Despite the abundance of charging options, all drivers tend to drive directly to their destinations without stopping to charge in later training episodes. This behaviour suggests that the availability of chargers does not influence drivers' decisions when their battery levels are already sufficient for short trips.

## Cluster Selection

The optimal number of clusters was determined using the elbow method. We calculate the inertia value for each cluster number, which is the sum of squared distances between data points and their cluster centroids. As shown in Figure S11-13, all the three scenarios have an elbow point when the cluster number is 3. As a result, during the cluster analysis, we choose to have three clusters in all the scenarios.

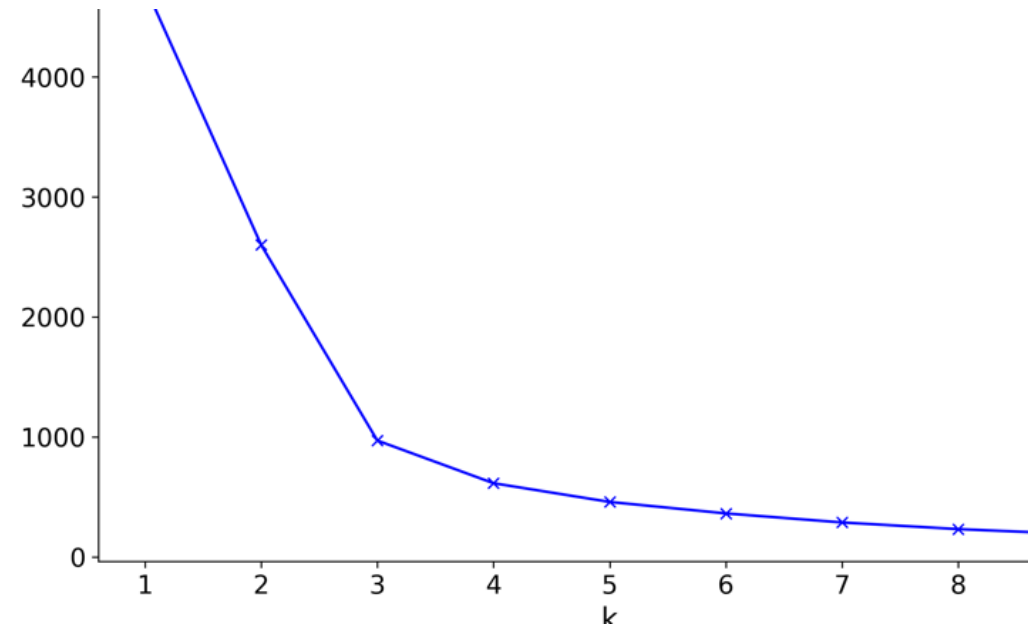

Figure S11 Inertia plot for buffer distance of 500m.

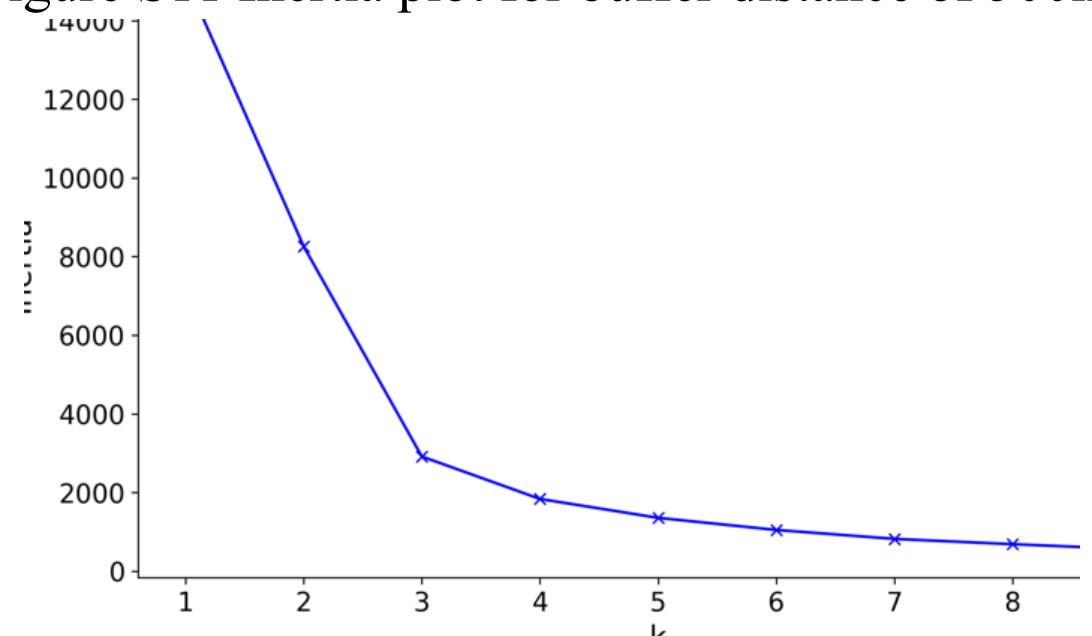

Figure S12 Inertia plot for buffer distance of 1000m.

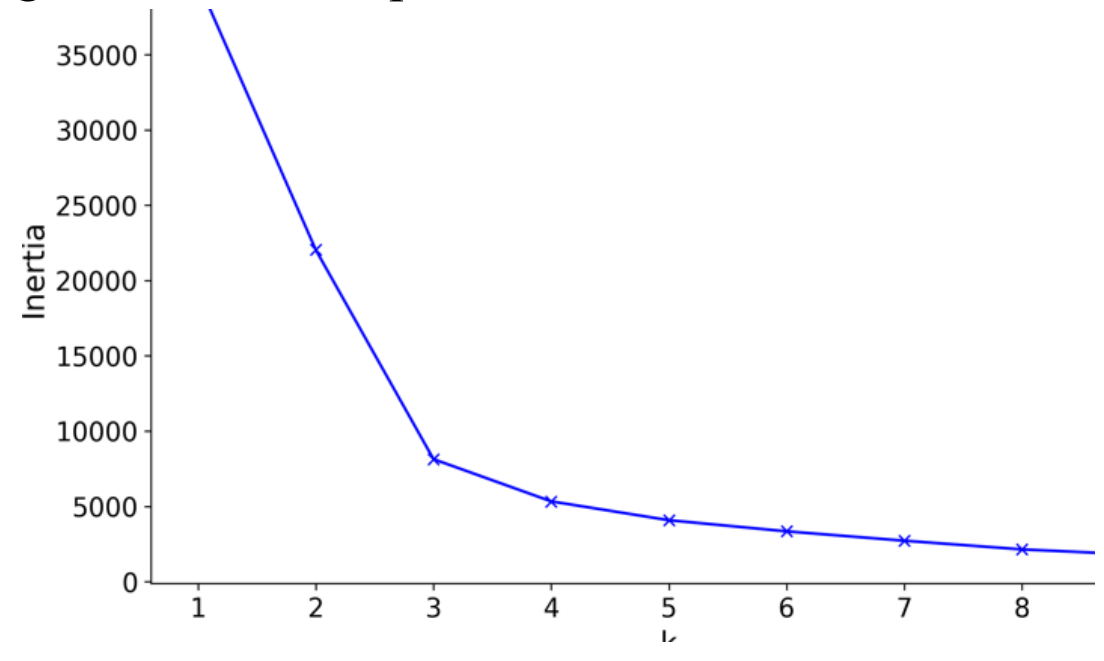

Figure S13 Inertia plot for buffer distance of 1500m.